\documentclass[reprint,twocolum,amsmath,amssymb,aps,prl,superscriptaddress, floatfix]{revtex4-1}

\usepackage[LGR,T1]{fontenc}
\usepackage{textgreek}
\usepackage{underscore}
\usepackage[utf8]{inputenc}
\usepackage{graphicx}
\usepackage{overpic}
 \usepackage{rotating}
 \usepackage{mwe, tikz}
 \usepackage{amsmath,amssymb}
\usepackage{pgfplots}
\usetikzlibrary{calc,arrows,decorations.markings,shapes.geometric} 
\usepackage{romannum}
\usepackage{xfp}
\usepackage{mathtools} 
\usepackage{color}
\usepackage[export]{adjustbox}
\usepackage{mwe, tikz}
\usepackage{rotating}
\usepackage{latexsym}
\usepackage{csquotes}
\usepackage{pict2e}
\usepackage{pgfplots}
\usepackage{float}
\usepackage{xr}
\usepackage{fp}
\usetikzlibrary{calc} 
\usepackage{bm}
\usepackage{float}
\usepackage{adjustbox}
\usepackage{booktabs, tabularx}
\usepackage{longtable}
\usepackage{hyperref}

\definecolor{light-gray}{gray}{0.85}

\newcommand{\drawSlope}[6]{ 
				\coordinate (center1) at (#1,#2); 
				 \FPeval{\nx}{cos(#4*pi/180)}%
				 \FPeval{\ny}{sin(#4*pi/180)}%
				 \FPeval{\absNx}{abs(\nx)}%
				 \FPeval{\absNy}{abs(\ny)}%
				\coordinate (b) at ($ (center1) + #3*(-\ny,+\nx) $);
				\coordinate (c) at ($ (center1) - #3*(-\ny,+\nx) $); 
				 \FPeval{\xx}{(-\nx*\ny)}%
				\ifdim\xx pt < 0pt 
				\coordinate (d) at ($ (c) -2*#3*\ny*(1,0) $);
				\node at ($(d) +(0,#3*\nx)-0.2*(\nx/\absNx,0)$) {{\color{#5}$#6$}}; 
				\node at ($(d)+(#3*\ny,0)-0.2*(0,\ny/\absNy)$) {{\color{#5}$\scriptstyle 1$}}; 
				\else
				\coordinate (d) at ($ (c) +2*#3*\nx*(0,1) $);
				\node at ($(d)-#3*\nx*(0,1)-0.2*\nx/\absNx*(1,0)$) {{\color{#5}$#6$}}; 
				\node at ($(d)-#3*\ny*(1,0)-0.2*\ny/\absNy*(0,1)$) {{\color{#5}$\scriptstyle 1$}}; 
				\fi
				\draw[#5,line width=0.1mm] (d) -- (b); 
				\draw[#5,line width=0.1mm] (d) -- (c); 
				\draw[#5,line width=0.1mm] (b) -- (c); 
} 

\definecolor{darkolivegreen}{rgb}{0.33, 0.42, 0.18}
\definecolor{darkspringgreen}{rgb}{0.09, 0.45, 0.27}
\definecolor{darkslategray}{rgb}{0.18, 0.31, 0.31}
\definecolor{darkred}{rgb}{0.55, 0.0, 0.0}

\newcommand{\etal}{~{\it et al.}}
\newcommand{\cf}{{\it cf.~}}

\begin{document}

\title{Dynamic nanoindentation and short-range order in equiatomic NiCoCr medium entropy alloy lead to novel density wave ordering}
\author{A. Naghdi}
\affiliation{%
NOMATEN Centre of Excellence, National Center for Nuclear Research, ul. A. Sołtana 7, 05-400 Swierk/Otwock, Poland
 }%
\author{F. J. Dom\'inguez-Guti\'errez}
\affiliation{%
NOMATEN Centre of Excellence, National Center for Nuclear Research, ul. A. Sołtana 7, 05-400 Swierk/Otwock, Poland
 }%
\author{W. Y. Huo}
\affiliation{%
NOMATEN Centre of Excellence, National Center for Nuclear Research, ul. A. Sołtana 7, 05-400 Swierk/Otwock, Poland
 }%
\affiliation{College of Mechanical and Electrical Engineering, Nanjing Forestry University, Nanjing, 210037, China}
\author{K. Karimi}
\affiliation{%
NOMATEN Centre of Excellence, National Center for Nuclear Research, ul. A. Sołtana 7, 05-400 Swierk/Otwock, Poland
 }%
\author{S. Papanikolaou}
\email{stefanos.papanikolaou@ncbj.gov.pl}
\affiliation{%
NOMATEN Centre of Excellence, National Center for Nuclear Research, ul. 
A. Sołtana 7, 05-400 Swierk/Otwock, Poland
 }%
\email{stefanos.papanikolaou@ncbj.gov.pl}

\begin{abstract}
Chemical short-range order (CSRO) is believed to be a key contributor to the exceptional properties of multicomponent alloys. However, direct validation and confirmation of CSRO has been highly elusive in most compounds. Recent studies for equiatomic NiCoCr alloys have shown that thermal treatments (i.e., annealing/aging) may facilitate and manipulate CSRO. In this work, by using molecular simulations, we show that nanomechanical probes, such as nanoindentation, may be utilized towards further manipulation of CSRO, providing explicit validation pathways. By using well established interatomic potentials, we perform hybrid Molecular-Dynamics/Monte-Carlo (MD/MC) at room temperature to demonstrate that particular dwell nanoindentation protocols can lead, through thermal MC equilibration, to the reorganization of CSRO under the indenter tip, to a density-wave stripe pattern (DWO). We characterize the novel DWO structures, that are directly correlated to incipient SRO but are highly anisotropic and dependent on local, nanoindentation-induced stress concentrations, and we show how they deeply originate from the peculiarities of the interatomic potentials. Furthermore, we show that the DWO patterns consistently scale up with the incipient plastic zone under the indenter tip, justifying the observation of the DWO patterning at any experimentally feasible nanoindentation depth. 

\end{abstract}

\maketitle

Concentrated multi--component alloys, and in particular, 
the celebrated Cantor 
alloys~\cite{CANTOR2004213,yeh2004nanostructured,miracle2014exploration}, 
such as equiatomic single--phase CoCrFeMnNi, have been 
instrumental into guiding the exploration for the discovery 
of affordable, durable alloys, suitable for applications under 
extreme conditions~\cite{li2019mechanical,shang2021mechanical}. 
It has been conjectured that a major contributor to the exceptional 
properties of these alloys is the formation of chemical 
short--range (1nm) order (CSRO) that may pin or/and obstruct
moving lattice defects, such as dislocations. 
While the observation of CSRO is quite common in such complex
alloys, its causal connection to \emph{exceptional} mechanical
properties has been a subject of intense debate. 
Extensive investigations have culminated towards the focus 
shining on the curious case of single--phase equiatomic NiCoCr. 
This alloy has outstanding mechanical properties, 
namely hardness, strength and ductility \cite{Gludovatz2016}, 
and there is plausible formation of CSROs in the alloys, 
especially after sample aging at high 
temperatures~\cite{zhang2020short,wu2021short,zhou2022atomic,chen2021direct,walsh2021magnetically, zhang2017local}. 
In addition, it has been observed that dislocation stacking 
fault widths are sensitive to the formation of such short--range 
order~\cite{zhang2020short}, which is commonly (in metallurgy)
associated with large(r) mechanical strength. 
Nevertheless, a deeper understanding of CSRO is required to
identify how to control and possibly, augment CSRO features,
and check its causal effects on mechanical properties. 
For this purpose, in this work, we theoretically consider
dwell nanoindentation as a possible way to locally manipulate
CSRO in equiatomic NiCoCr, and lead to causal connections
between hardness and microscopy--resolvable 
nanostructural features. 
We utilize molecular simulation, as well as Monte Carlo methods, 
and we show that CSRO (if it exists) will be unstable to the
formation of unconventional density-wave ordered stripe 
patterns (DWO), that are highly anisotropic and originate 
due to interatomic potential features.

CSRO is commonly observable at the nanoscale, ranging between 
$0.5$--$2$nm in linear size~\cite{zhou2022atomic}. 
Starting from a random solid solution (RSS), it is natural to 
expect that constituent element enthalpic interactions may cause
CSRO formation at the atomic scale 
\cite{Ding2019,zhang2020short,Widom2014, TAMM2015307, Santodonato2015, singh, zhang2017local, koch, Fernandez-Caballero2017, MA201864, Li2019, Oh2019, JIAN2020352, Yin2020, wu2021short}. 
CSROs have a notable effect on defects (dislocations, 
interstitials and vacancies etc.) and their dynamical behaviour, 
as well as macroscopic alloy mechanical properties  
\cite{Ding2019, zhang2020short, Li2019, Oh2019, JIAN2020352, Yin2020, wu2021short}. 
CSRO is commonly inferred through implicit experimental 
measurements, through their relation to macroscopic properties, 
such as stacking--fault energy, hardness,  irradiation 
effects, etc. 
\cite{koch, Fernandez-Caballero2017, MA201864, GEORGE2020435}. 
CSRO is also investigated in theoretical and computational works
\cite{Li2019, walsh2021magnetically, Ding}, with the most intense
focus being on the case of equiatomic NiCoCr \cite{Li2019}. 
Molecular Dynamics (MD), spin polarized Density Functional
Theory (DFT)   \cite{walsh2021magnetically}, but also 
Monte Carlo (MC) simulations have been utilized to 
characterize CSRO structural features~\cite{Ding}.

Creep deformation studies in multi--component alloys 
are abundant~\cite{Lenz2019TensionCompressionAO, Bezold2020, XUE2018129, cryst10111058, Rhein2018, FENG201899, may2005strain}. 
Here, we aim at utilizing creep deformation to comprehend in 
a deeper sense the character of CSRO by promoting 
non--trivial predictions for elemental density wave ordering 
at the nanoscale. 
While dynamical uniaxial testing~\cite{may2005strain, Chinh, Li2007, WEI200471, Hoppel, ni2017probing} and nanoindentation, alongside microscopy, have been 
used for the elucidation of lattice defect deformation 
mechanisms~\cite{Maier2013}, the constant--load dwell 
nanoindentation tests, between $1$ minute to $10$ hours of 
dwell time, have been solely popular for investigating the 
relation between hardness and indentation strain--rate over a, 
possibly wide, temperature 
range~\cite{Poisl1995, Lucas1999, Stone2010, Choi2012}. 
We use this concept to generate predictions for CSRO 
nano--patterning at room temperature that may be directly 
observable using electron microscopy techniques. 
While this work is fully focused on modeling aspects and 
predictions for CSRO saturation regimes that may emerge at 
each loading depth, prior alloy studies~\cite{Poisl1995, Lucas1999, Stone2010, Choi2012} shall 
allow us to conclude that the proposed scenario is attainable 
at room conditions.

\begin{figure}[!t]
    \centering
    \includegraphics[width=8cm]{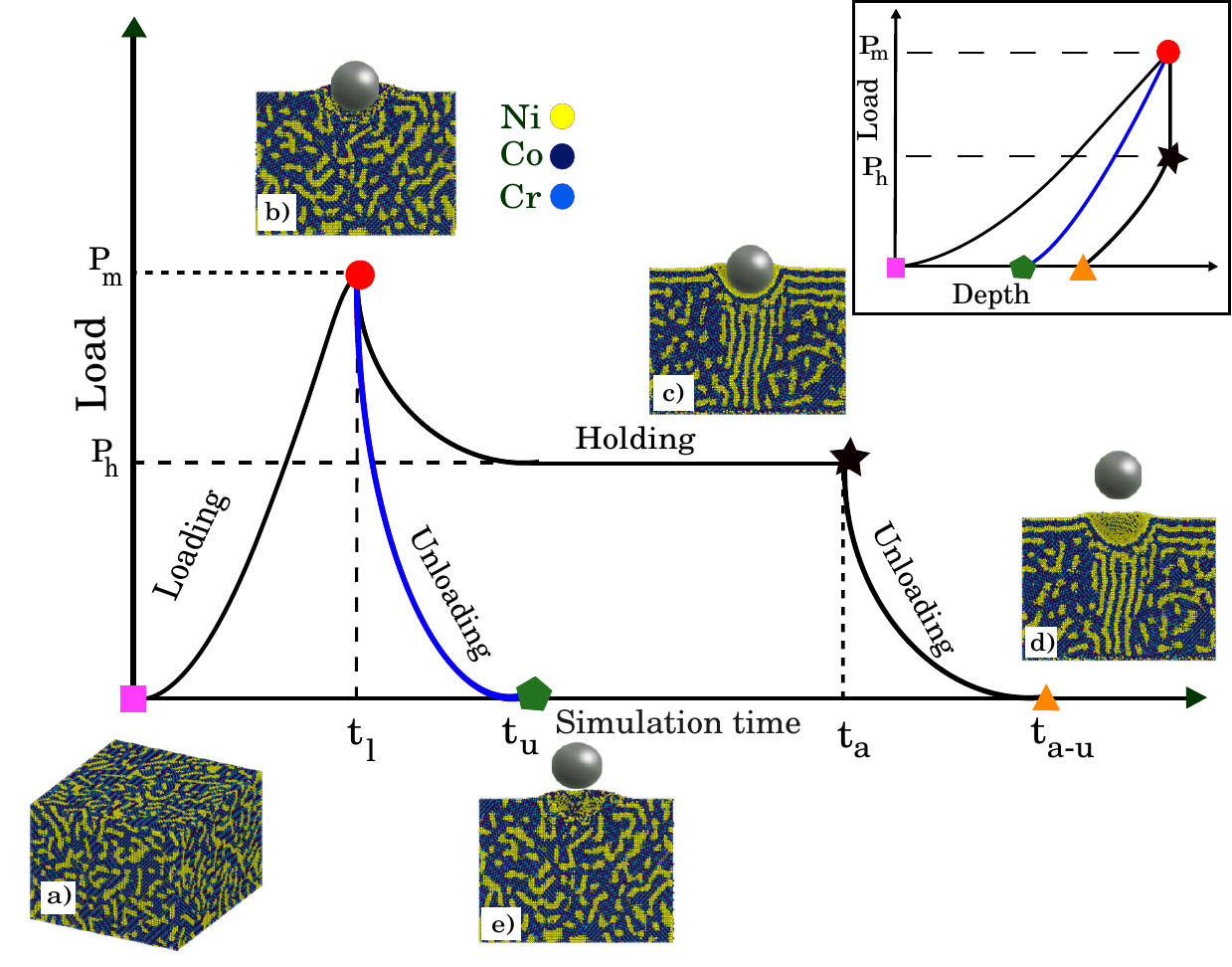}
    \caption{Nanoindentation protocol for Short-Range Order re-organization in equiatomic NiCoCr alloys. The process starts when the aged NiCoCr sample \textbf{(a)} is indented up to a certain depth \textbf{(b)}. Afterwards, the indenter's velocity is set to zero, this is the moment when the Hybrid MD-MC process starts again and leads to a configuration in which Ni and Co-Cr segregations or no longer randomly distributed but are reorganized and form stripe patterns under the indenter tip \textbf{(c)}. The resulting pattern retains its shape even after the indenter is removed from the sample \textbf{(d)}. This pattern is not observed during normal loading-unloading nanoindentation simulation like the process shown by the blue line \textbf{(e)}.} 
    
    \label{fig:Process}
\end{figure}

In this paper, we utilize hybrid MC-MD simulations, using 
LAMMPS \cite{LAMMPS}, to demonstrate the plausible 
thermomechanical effects of a dwell nanoindentation scenario
in single--phase equiatomic RSS NiCoCr. 
RSS samples were generated using random elemental sampling
on appropriate face--centered cubic (FCC) lattices, with 
crystal orientations of $x=[100]$, $y=[010]$ and $z=[001]$ 
or $x=[100]$, $y=[01\bar1]$ and $z=[011]$, and dimensions 
$25.85 \times23.59 \times15.14$nm$^3$. Samples then undergo 
energy minimization at $T=0K$ and then relaxation  for $100 ps$ 
at $T=300K$ with  time discretization  $\Delta t\simeq 1.0$fs, 
in an NPT ensemble with a Nose--Hoover thermostat and barostat 
with relaxation time scales of $\tau_d^\text{therm}=10$ fs and 
$\tau_d^\text{bar}=100$ fs, using an EAM-based potential, firstly 
proposed by Li\etal~\cite{Li2019}, who also showed that 
annealing of the samples (even at room temperature) leads to the 
formation of characteristic Ni--rich SRO patterning, which we 
will generally refer to as SRO-samples in this paper.  
The nanoindentation process was performed through an NVE ensemble. 
Furthermore, nanoindentation was performed along $z$, using a tip 
at radii of $3.5$, $5$ and $7$nm. 
Periodic boundary conditions (BC) were implemented along $x$ 
and $y$,  while along $z$, BC was fixed at the bottom boundary, 
and free at the top. 
Moreover, a $3$\AA--thick layer of atoms was frozen at the 
bottom. The indenter tip is assumed to be a rigid sphere with 
force:  \(\ F(t) = K(\overrightarrow{\text{r}}(t)-R)^2\), 
where $K = 1000$ eV/\AA$^3$ and $R$ is the tip radius, moving 
along z direction with a speed of $v = 20 m/s$~\cite{DOMINGUEZGUTIERREZ2021141912,KURPASKA2022110639}. 
As shown in Fig.~\ref{fig:Process}, when the indenter reached 
the target depth ($1$nm, $2$nm or $3$nm), $v$ was set to zero
and then MC thermal relaxation ('holding') process was 
performed using the variance-constrained semi--grand 
canonical (VCSGC) ensemble \cite{PhysRevB.85.184203,Li2019} 
at $T=300K$. 
The chemical potential differences $\Delta\mu_\text{Ni-Cr} 
= -0.31$ and $\Delta\mu_\text{Ni-Co} = 0.021$ and the variance 
constraint $\kappa=1000$ are set as in Ref.~\cite{Li2019}. 
The 'Holding' part (\cf Fig.~\ref{fig:Process}) includes $1$ MC 
cycle for every $20$ MD steps within the VCSGC ensemble for 
a total number of $150,000$ MC cycles, ensuring that the 
thermalized configuration contains stable SRO patterning.

In the studied protocol (\cf Fig.~\ref{fig:Process}), equiatomic
NiCoCr, which has been shown to display Ni--rich SRO 
patterns~\cite{Li2019}, is indented to a specified indentation 
depth in the 'Loading' step. 
The CSRO changes only during the 'Holding' step, where the indenter
is held fixed and thermomechanical MC--MD relaxation is performed
at room temperature (\cf Fig.~\ref{fig:Process}). 
Due to 'Holding', a load drop is commonly observed from 
maximum $P_m$ at time $t_l$ to $P_h$ at time $t_a$. 
The 'Unloading' step consists in removing the load altogether 
at a velocity of $v=-20$nm/s, using MD, and then perform further
MC--MD relaxation at the sample. 
We find that a characteristic DWO pattern emerges after 
'Unloading' at $t_\text{a-u}$ (\cf Fig.~\ref{fig:Process}), 
that would not appear without the 'Holding' stage. 
The whole protocol is also illustrated in the Load--Depth plot
in the inset of Fig.~\ref{fig:Process}.

\begin{figure*}[t]
   \centering
    \includegraphics[width=18cm]{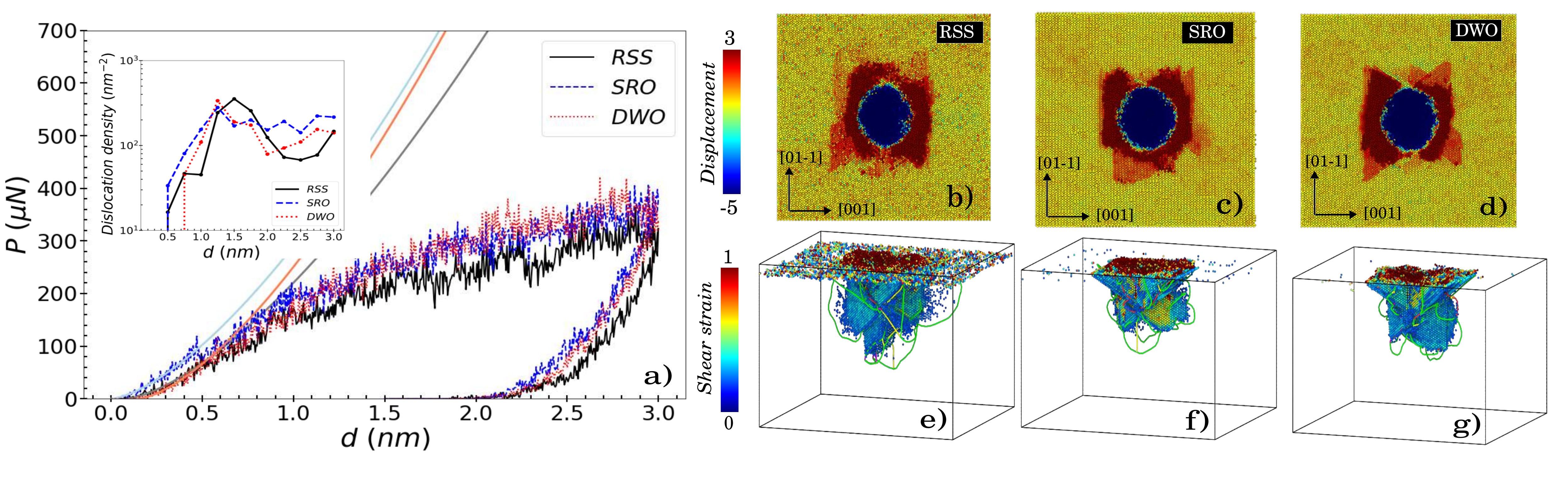}
    \caption{
    Characterization of nanoindentation-driven reorganization of Short-Range Ordering. The nanoindentation LD curves for a RSS, an annealed sample and a sample with stripe patterns are shown in \textbf{(a)}. Obviously, the annealed and stripy samples are stronger. \textbf{(b-c)} are the pile up patterns for RSS, annealed and the sample with stripes. The strength of the segregated samples are also depicted in \textbf{(e-f)} (RSS, annealed and striped sample, respectively.) in the sense that the plastic region is smaller for the two latter ones.}
   \label{fig:characterization}
\end{figure*}

{The characterization of the resulting nanoindentation--driven 
DWO pattern observed in our simulations, is shown and compared
to a RSS and a Ni--rich SRO sample in Fig~\ref{fig:characterization}. 
The DWO and SRO rich samples happen to have a larger mechanical 
strength and hardness than a RSS, as their load-depth (P--d) 
curves illustrate in Fig~\ref{fig:characterization}(a). 
This could be due to the already observed Ni--dominated solute 
segregation, that is pinning and obstructing dislocation 
motion~\cite{esfandiarpour2022edge,naghdi2022dislocation}. 
This  phenomenon may also be implied by a drastic dislocation
density increase for DWO and SRO samples at smaller depths, 
compared to RSS ones, as shown in the inset of 
Fig~\ref{fig:characterization}(a). 
Additionally, the RSS plastic zone is typically larger than those of the DWO and SRO samples, as shown in 
Fig~\ref{fig:characterization}(e-f), concluding that SRO 
and DWO have analogous dislocation pinning effects.} 

\begin{figure}[!b]
   \centering
   \includegraphics[width=8cm]{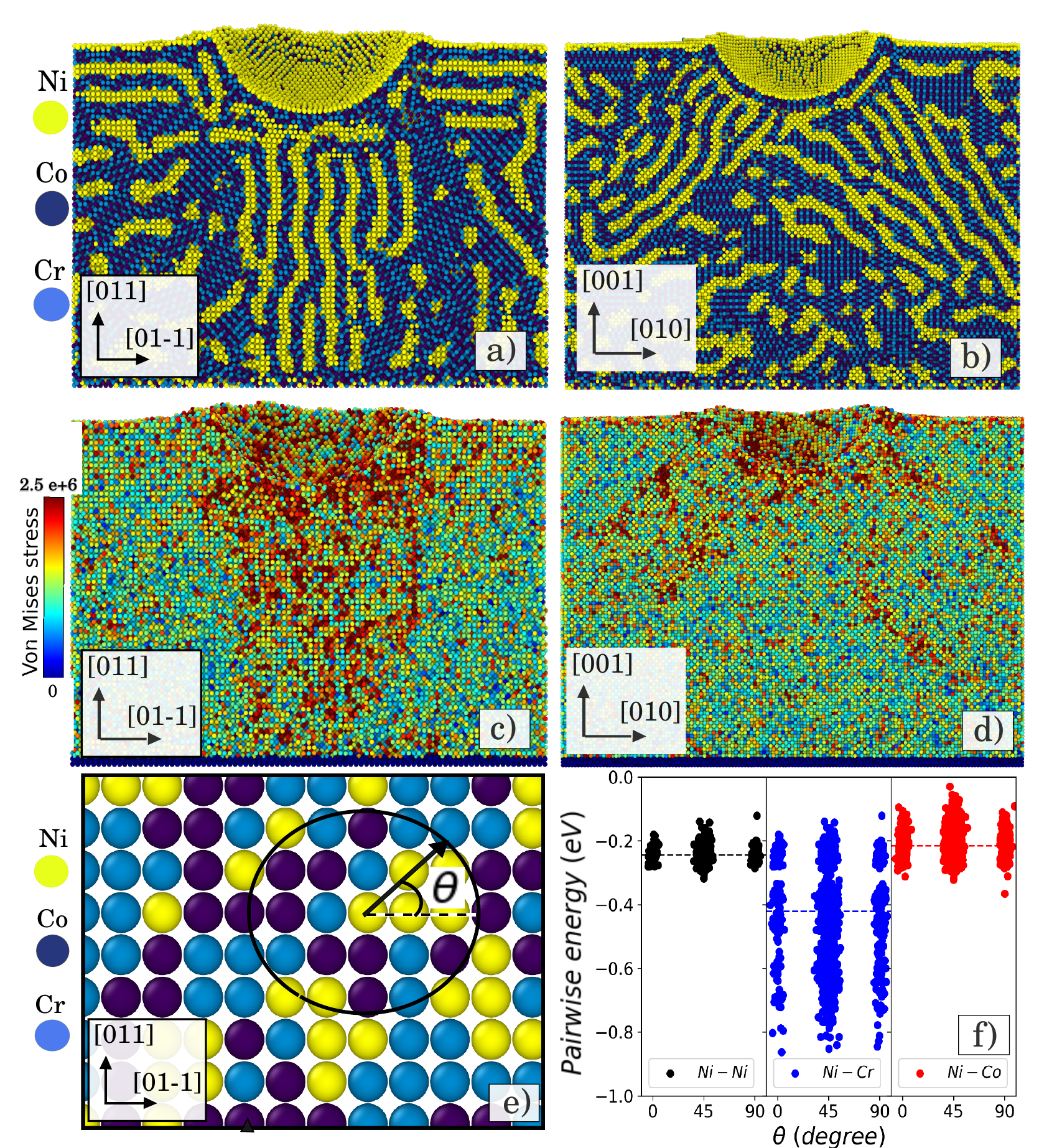}
   \caption{
   Origin of nanoindentation-driven reorganization of Short-Range Ordering. \textbf{(a-b)} are the SRO patterns found along the $z[011]$ and $z[001]$ orientations, respectively. \textbf{(c-d)} illustrate the correlation between the stripe patterns' orientation shown in \textbf{(a-b)} and the Von-Mises stress. The pairwise energy between Ni-Ni, Ni-Cr and Ni-Co atoms within each atom's neighbour list \textbf{(e)} is plotted in the same plane in \textbf{(a)} (but for separate Ni, Ni-Cr and Ni-Co crystals) with respect to the in-plane angle of the pair atoms \textbf{(\cf F)}. As shown in \textbf{(\cf F)}, the Ni-Cr pairs have an average energy lower by 3 orders of magnitude compared to Ni-Ni and Ni-Co pairs.} 
   \label{fig:Orientations}
\end{figure}

The character and origin of the emergent DWO onset was further investigated, by considering the effect of crystal orientation. We find that the behavior for crystals along $x=[100]$, $y=[010]$ and $z=[001]$ (\cf Fig.~\ref{fig:Orientations}(b)) is drastically different than the original orientation behavior (\cf Fig.~\ref{fig:Orientations}(a)) with the DWO orientation being tilted. However, further inspection shows that DWO forms along $z=[011]$ universally in the same direction relative to the crystal (\cf Fig.~\ref{fig:Orientations}(b)). The DWO pattern also aligns with the maximum Von-Mises stress, accumulated in the $[011]$ planes (\cf Fig.~\ref{fig:Orientations}.(c-d)), a fact that implies material anisotropy. To further elucidate this issue, we estimate the pairwise potential energy for the Ni-Ni, Ni-Cr and Ni-Co pairs in a RSS crystal oriented in $x=[100]$, $y=[01\bar1]$ and $z=[011]$, as shown in Fig~\ref{fig:Orientations}.(\cf F), is calculated for each atom (\cf Fig.~\ref{fig:Orientations}.(e)), considering only the first nearest neighbours defined by the first pick of the pair correlation function $g(r)$ ($3$\AA). With this input, we compare the total energy of a DWO ansatz (see Supplementary Material for details) order that compares well with Fig.~\ref{fig:Orientations}(a) and contains the inter-stripe distance as a free parameter. By comparing this DWO ansatz energy to the energy of RSS samples, we find that the optimal inter-stripe distance is very close to the one realized in MD simulations. In this way, we conclude that the DWO emergence is deeply linked to the energetic features of the interatomic potential, which is also the key cause of SRO emergence in equiatomic NiCoCr simulations.

\begin{figure*}[t]
   \centering
   \includegraphics[width=18cm]{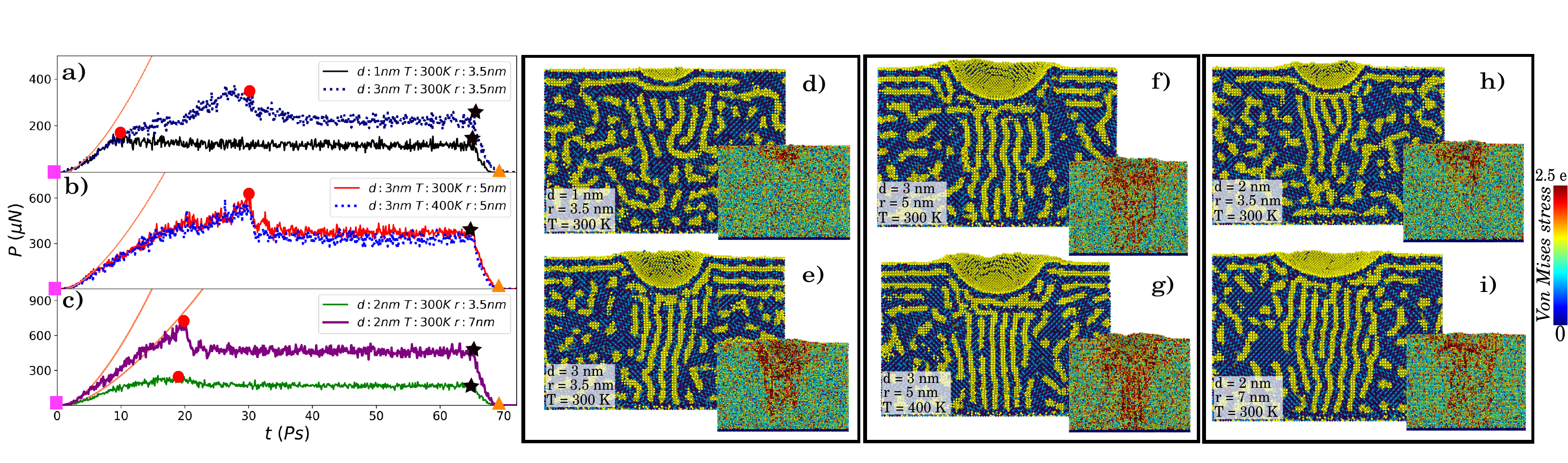}
   \caption{Size effects of nanoindentation-driven reorganization of Short-Range Ordering. Figure \textbf{(a)} shows the load drop is bigger for bigger holding-depths, which results in more pronounced stripe patterns \textbf{(b)}. Also, the same correlation exists between the stripe patterns and the Von-Mises stress values (shown in inset snapshots of \textbf{(b-c)}). Following this, \textbf{(c)} shows by increasing the temperature by $100K$ the Von-Mises stress increases (more red spots in the inset of figure \textbf{(c)}) and leads to a more organized stripe pattern, although the indenter's depth and radius is the same for the two cases. As it is expected, a bigger indenter radius induces a bigger stress in the sample and leads to a bigger pattern as it is shown in \textbf{(d)}.} 
   \label{fig:Size}
\end{figure*}

{The emergent DWO displays strong size effects~\cite{papanikolaou2017avalanches}, which are dependent on the indentation depth and indenter tip radius as a function of temperature (\cf Fig.~\ref{fig:Size}). In our displacement-controlled tests, we find that load-time (P-t) curves display a larger load drop~\cite{papanikolaou2017avalanches} during 'Holding' as depth or tip radius increases, leading to spatially extended DWO (\cf Fig.~\ref{fig:Size}), resembling the plastic zone size. While not studied here, we also expect that size-dependent strain bursts should be observed in load-controlled tests. More specifically, the protocol discussed in Fig.~\ref{fig:Process} is implemented for two different indenter depth ($1$ and $3nm$) while the temperature ($300K$) and the indenter tip radii ($3.5nm$) are kept fixed. Furthermore, increasing the indenter radii from $2$ nm to $7$ nm gives rise to a larger plastic zone and as a result, a larger DWO pattern. This effect can also be observed from the larger load drop observed in the P-t curves for the larger radii (\cf Fig.~\ref{fig:Size}(c)). However, increasing the temperature (from $300K$ to $400K$), while the indenter depth and radii are the same, also results in a more organized DWO pattern shown in Fig.~\ref{fig:Size}(\cf F-g), but there is no pronounced size effect, as shown by the absence of a load-drop decrease (\cf Fig.~\ref{fig:Size}.(b)). }
 
{Given the elusive character of SRO formation in advanced alloys~\cite{GEORGE2020435}, the described dynamic nanoindentation protocol appears to be a plausible candidate for nano-scale manipulation and control of CSRO patterns in equiatomic NiCoCr and possibly, other multicomponent alloys. By the investigation of thermomechanical features and size effects, we conclude that atomic scale Ni-rich segregation strongly influences the mechanical properties of equiatomic NiCoCr, in a way that can be quantified in dynamic nanoindentation. CSRO reorganization mechanisms in NiCoCr, are shown to be energetic in character (as opposed to entropic) and are highly anisotropic. The origin of DWO emergence is tracked back at the potential energy surface of a RSS crystal. 
Finally, the pronounced observed size effects of the emergent DWO, suggest that, for experimentally relevant nanoindentation depths and tip radii, the emergent DWO shall be visible under common electron microscopy tools at the nanoscale.
}

\begin{acknowledgements}
{\it Acknowledgements -- }
We would like to thank Mikko Alava, Pawel Sobkowicz and Lukasz Kurpaska for fruitful discussions on features of short-range order. We acknowledge support from the European Union Horizon 2020 research
 and innovation program under grant agreement no. 857470 and from the 
 European Regional Development Fund via the Foundation for Polish 
 Science International Research Agenda PLUS program grant 
 No. MAB PLUS/2018/8.
 \end{acknowledgements}


\begin{thebibliography}{50}%
\makeatletter
\providecommand \@ifxundefined [1]{%
 \@ifx{#1\undefined}
}%
\providecommand \@ifnum [1]{%
 \ifnum #1\expandafter \@firstoftwo
 \else \expandafter \@secondoftwo
 \fi
}%
\providecommand \@ifx [1]{%
 \ifx #1\expandafter \@firstoftwo
 \else \expandafter \@secondoftwo
 \fi
}%
\providecommand \natexlab [1]{#1}%
\providecommand \enquote  [1]{``#1''}%
\providecommand \bibnamefont  [1]{#1}%
\providecommand \bibfnamefont [1]{#1}%
\providecommand \citenamefont [1]{#1}%
\providecommand \href@noop [0]{\@secondoftwo}%
\providecommand \href [0]{\begingroup \@sanitize@url \@href}%
\providecommand \@href[1]{\@@startlink{#1}\@@href}%
\providecommand \@@href[1]{\endgroup#1\@@endlink}%
\providecommand \@sanitize@url [0]{\catcode `\\12\catcode `\$12\catcode
  `\&12\catcode `\#12\catcode `\^12\catcode `\_12\catcode `\%12\relax}%
\providecommand \@@startlink[1]{}%
\providecommand \@@endlink[0]{}%
\providecommand \url  [0]{\begingroup\@sanitize@url \@url }%
\providecommand \@url [1]{\endgroup\@href {#1}{\urlprefix }}%
\providecommand \urlprefix  [0]{URL }%
\providecommand \Eprint [0]{\href }%
\providecommand \doibase [0]{http://dx.doi.org/}%
\providecommand \selectlanguage [0]{\@gobble}%
\providecommand \bibinfo  [0]{\@secondoftwo}%
\providecommand \bibfield  [0]{\@secondoftwo}%
\providecommand \translation [1]{[#1]}%
\providecommand \BibitemOpen [0]{}%
\providecommand \bibitemStop [0]{}%
\providecommand \bibitemNoStop [0]{.\EOS\space}%
\providecommand \EOS [0]{\spacefactor3000\relax}%
\providecommand \BibitemShut  [1]{\csname bibitem#1\endcsname}%
\let\auto@bib@innerbib\@empty
\bibitem [{\citenamefont {Cantor}\ \emph {et~al.}(2004)\citenamefont {Cantor},
  \citenamefont {Chang}, \citenamefont {Knight},\ and\ \citenamefont
  {Vincent}}]{CANTOR2004213}%
  \BibitemOpen
  \bibfield  {author} {\bibinfo {author} {\bibfnamefont {B.}~\bibnamefont
  {Cantor}}, \bibinfo {author} {\bibfnamefont {I.}~\bibnamefont {Chang}},
  \bibinfo {author} {\bibfnamefont {P.}~\bibnamefont {Knight}}, \ and\ \bibinfo
  {author} {\bibfnamefont {A.}~\bibnamefont {Vincent}},\ }\href {\doibase
  https://doi.org/10.1016/j.msea.2003.10.257} {\bibfield  {journal} {\bibinfo
  {journal} {Materials Science and Engineering: A}\ }\textbf {\bibinfo {volume}
  {375-377}},\ \bibinfo {pages} {213} (\bibinfo {year} {2004})}\BibitemShut
  {NoStop}%
\bibitem [{\citenamefont {Yeh}\ \emph {et~al.}(2004)\citenamefont {Yeh},
  \citenamefont {Chen}, \citenamefont {Lin}, \citenamefont {Gan}, \citenamefont
  {Chin}, \citenamefont {Shun}, \citenamefont {Tsau},\ and\ \citenamefont
  {Chang}}]{yeh2004nanostructured}%
  \BibitemOpen
  \bibfield  {author} {\bibinfo {author} {\bibfnamefont {J.-W.}\ \bibnamefont
  {Yeh}}, \bibinfo {author} {\bibfnamefont {S.-K.}\ \bibnamefont {Chen}},
  \bibinfo {author} {\bibfnamefont {S.-J.}\ \bibnamefont {Lin}}, \bibinfo
  {author} {\bibfnamefont {J.-Y.}\ \bibnamefont {Gan}}, \bibinfo {author}
  {\bibfnamefont {T.-S.}\ \bibnamefont {Chin}}, \bibinfo {author}
  {\bibfnamefont {T.-T.}\ \bibnamefont {Shun}}, \bibinfo {author}
  {\bibfnamefont {C.-H.}\ \bibnamefont {Tsau}}, \ and\ \bibinfo {author}
  {\bibfnamefont {S.-Y.}\ \bibnamefont {Chang}},\ }\href@noop {} {\bibfield
  {journal} {\bibinfo  {journal} {Advanced engineering materials}\ }\textbf
  {\bibinfo {volume} {6}},\ \bibinfo {pages} {299} (\bibinfo {year}
  {2004})}\BibitemShut {NoStop}%
\bibitem [{\citenamefont {Miracle}\ \emph {et~al.}(2014)\citenamefont
  {Miracle}, \citenamefont {Miller}, \citenamefont {Senkov}, \citenamefont
  {Woodward}, \citenamefont {Uchic},\ and\ \citenamefont
  {Tiley}}]{miracle2014exploration}%
  \BibitemOpen
  \bibfield  {author} {\bibinfo {author} {\bibfnamefont {D.~B.}\ \bibnamefont
  {Miracle}}, \bibinfo {author} {\bibfnamefont {J.~D.}\ \bibnamefont {Miller}},
  \bibinfo {author} {\bibfnamefont {O.~N.}\ \bibnamefont {Senkov}}, \bibinfo
  {author} {\bibfnamefont {C.}~\bibnamefont {Woodward}}, \bibinfo {author}
  {\bibfnamefont {M.~D.}\ \bibnamefont {Uchic}}, \ and\ \bibinfo {author}
  {\bibfnamefont {J.}~\bibnamefont {Tiley}},\ }\href@noop {} {\bibfield
  {journal} {\bibinfo  {journal} {Entropy}\ }\textbf {\bibinfo {volume} {16}},\
  \bibinfo {pages} {494} (\bibinfo {year} {2014})}\BibitemShut {NoStop}%
\bibitem [{\citenamefont {Li}\ \emph {et~al.}(2019{\natexlab{a}})\citenamefont
  {Li}, \citenamefont {Zhao}, \citenamefont {Ritchie},\ and\ \citenamefont
  {Meyers}}]{li2019mechanical}%
  \BibitemOpen
  \bibfield  {author} {\bibinfo {author} {\bibfnamefont {Z.}~\bibnamefont
  {Li}}, \bibinfo {author} {\bibfnamefont {S.}~\bibnamefont {Zhao}}, \bibinfo
  {author} {\bibfnamefont {R.~O.}\ \bibnamefont {Ritchie}}, \ and\ \bibinfo
  {author} {\bibfnamefont {M.~A.}\ \bibnamefont {Meyers}},\ }\href@noop {}
  {\bibfield  {journal} {\bibinfo  {journal} {Progress in Materials Science}\
  }\textbf {\bibinfo {volume} {102}},\ \bibinfo {pages} {296} (\bibinfo {year}
  {2019}{\natexlab{a}})}\BibitemShut {NoStop}%
\bibitem [{\citenamefont {Shang}\ \emph {et~al.}(2021)\citenamefont {Shang},
  \citenamefont {Brechtl}, \citenamefont {Psitidda},\ and\ \citenamefont
  {Liaw}}]{shang2021mechanical}%
  \BibitemOpen
  \bibfield  {author} {\bibinfo {author} {\bibfnamefont {Y.}~\bibnamefont
  {Shang}}, \bibinfo {author} {\bibfnamefont {J.}~\bibnamefont {Brechtl}},
  \bibinfo {author} {\bibfnamefont {C.}~\bibnamefont {Psitidda}}, \ and\
  \bibinfo {author} {\bibfnamefont {P.~K.}\ \bibnamefont {Liaw}},\ }\href@noop
  {} {\bibfield  {journal} {\bibinfo  {journal} {arXiv preprint
  arXiv:2102.09055}\ } (\bibinfo {year} {2021})}\BibitemShut {NoStop}%
\bibitem [{\citenamefont {Gludovatz}\ \emph {et~al.}(2016)\citenamefont
  {Gludovatz}, \citenamefont {Hohenwarter}, \citenamefont {Thurston},
  \citenamefont {Bei}, \citenamefont {Wu}, \citenamefont {George},\ and\
  \citenamefont {Ritchie}}]{Gludovatz2016}%
  \BibitemOpen
  \bibfield  {author} {\bibinfo {author} {\bibfnamefont {B.}~\bibnamefont
  {Gludovatz}}, \bibinfo {author} {\bibfnamefont {A.}~\bibnamefont
  {Hohenwarter}}, \bibinfo {author} {\bibfnamefont {K.~V.~S.}\ \bibnamefont
  {Thurston}}, \bibinfo {author} {\bibfnamefont {H.}~\bibnamefont {Bei}},
  \bibinfo {author} {\bibfnamefont {Z.}~\bibnamefont {Wu}}, \bibinfo {author}
  {\bibfnamefont {E.~P.}\ \bibnamefont {George}}, \ and\ \bibinfo {author}
  {\bibfnamefont {R.~O.}\ \bibnamefont {Ritchie}},\ }\href@noop {} {\bibfield
  {journal} {\bibinfo  {journal} {Nature Communications}\ }\textbf {\bibinfo
  {volume} {7}},\ \bibinfo {pages} {10602} (\bibinfo {year}
  {2016})}\BibitemShut {NoStop}%
\bibitem [{\citenamefont {Zhang}\ \emph {et~al.}(2020)\citenamefont {Zhang},
  \citenamefont {Zhao}, \citenamefont {Ding}, \citenamefont {Chong},
  \citenamefont {Jia}, \citenamefont {Ophus}, \citenamefont {Asta},
  \citenamefont {Ritchie},\ and\ \citenamefont {Minor}}]{zhang2020short}%
  \BibitemOpen
  \bibfield  {author} {\bibinfo {author} {\bibfnamefont {R.}~\bibnamefont
  {Zhang}}, \bibinfo {author} {\bibfnamefont {S.}~\bibnamefont {Zhao}},
  \bibinfo {author} {\bibfnamefont {J.}~\bibnamefont {Ding}}, \bibinfo {author}
  {\bibfnamefont {Y.}~\bibnamefont {Chong}}, \bibinfo {author} {\bibfnamefont
  {T.}~\bibnamefont {Jia}}, \bibinfo {author} {\bibfnamefont {C.}~\bibnamefont
  {Ophus}}, \bibinfo {author} {\bibfnamefont {M.}~\bibnamefont {Asta}},
  \bibinfo {author} {\bibfnamefont {R.~O.}\ \bibnamefont {Ritchie}}, \ and\
  \bibinfo {author} {\bibfnamefont {A.~M.}\ \bibnamefont {Minor}},\ }\href@noop
  {} {\bibfield  {journal} {\bibinfo  {journal} {Nature}\ }\textbf {\bibinfo
  {volume} {581}},\ \bibinfo {pages} {283} (\bibinfo {year}
  {2020})}\BibitemShut {NoStop}%
\bibitem [{\citenamefont {Wu}\ \emph {et~al.}(2021)\citenamefont {Wu},
  \citenamefont {Zhang}, \citenamefont {Yuan}, \citenamefont {Huang},
  \citenamefont {Wen}, \citenamefont {Wang}, \citenamefont {Zhang},
  \citenamefont {Wu}, \citenamefont {Liu}, \citenamefont {Wang} \emph
  {et~al.}}]{wu2021short}%
  \BibitemOpen
  \bibfield  {author} {\bibinfo {author} {\bibfnamefont {Y.}~\bibnamefont
  {Wu}}, \bibinfo {author} {\bibfnamefont {F.}~\bibnamefont {Zhang}}, \bibinfo
  {author} {\bibfnamefont {X.}~\bibnamefont {Yuan}}, \bibinfo {author}
  {\bibfnamefont {H.}~\bibnamefont {Huang}}, \bibinfo {author} {\bibfnamefont
  {X.}~\bibnamefont {Wen}}, \bibinfo {author} {\bibfnamefont {Y.}~\bibnamefont
  {Wang}}, \bibinfo {author} {\bibfnamefont {M.}~\bibnamefont {Zhang}},
  \bibinfo {author} {\bibfnamefont {H.}~\bibnamefont {Wu}}, \bibinfo {author}
  {\bibfnamefont {X.}~\bibnamefont {Liu}}, \bibinfo {author} {\bibfnamefont
  {H.}~\bibnamefont {Wang}},  \emph {et~al.},\ }\href@noop {} {\bibfield
  {journal} {\bibinfo  {journal} {Journal of Materials Science and Technology}\
  }\textbf {\bibinfo {volume} {62}},\ \bibinfo {pages} {214} (\bibinfo {year}
  {2021})}\BibitemShut {NoStop}%
\bibitem [{\citenamefont {Zhou}\ \emph {et~al.}(2022)\citenamefont {Zhou},
  \citenamefont {Wang}, \citenamefont {Wang}, \citenamefont {Chen},
  \citenamefont {Jiang}, \citenamefont {Zhou}, \citenamefont {Yuan},
  \citenamefont {Wu}, \citenamefont {Cheng},\ and\ \citenamefont
  {Ma}}]{zhou2022atomic}%
  \BibitemOpen
  \bibfield  {author} {\bibinfo {author} {\bibfnamefont {L.}~\bibnamefont
  {Zhou}}, \bibinfo {author} {\bibfnamefont {Q.}~\bibnamefont {Wang}}, \bibinfo
  {author} {\bibfnamefont {J.}~\bibnamefont {Wang}}, \bibinfo {author}
  {\bibfnamefont {X.}~\bibnamefont {Chen}}, \bibinfo {author} {\bibfnamefont
  {P.}~\bibnamefont {Jiang}}, \bibinfo {author} {\bibfnamefont
  {H.}~\bibnamefont {Zhou}}, \bibinfo {author} {\bibfnamefont {F.}~\bibnamefont
  {Yuan}}, \bibinfo {author} {\bibfnamefont {X.}~\bibnamefont {Wu}}, \bibinfo
  {author} {\bibfnamefont {Z.}~\bibnamefont {Cheng}}, \ and\ \bibinfo {author}
  {\bibfnamefont {E.}~\bibnamefont {Ma}},\ }\href@noop {} {\bibfield  {journal}
  {\bibinfo  {journal} {Acta Materialia}\ }\textbf {\bibinfo {volume} {224}},\
  \bibinfo {pages} {117490} (\bibinfo {year} {2022})}\BibitemShut {NoStop}%
\bibitem [{\citenamefont {Chen}\ \emph {et~al.}(2021)\citenamefont {Chen},
  \citenamefont {Wang}, \citenamefont {Cheng}, \citenamefont {Zhu},
  \citenamefont {Zhou}, \citenamefont {Jiang}, \citenamefont {Zhou},
  \citenamefont {Xue}, \citenamefont {Yuan}, \citenamefont {Zhu} \emph
  {et~al.}}]{chen2021direct}%
  \BibitemOpen
  \bibfield  {author} {\bibinfo {author} {\bibfnamefont {X.}~\bibnamefont
  {Chen}}, \bibinfo {author} {\bibfnamefont {Q.}~\bibnamefont {Wang}}, \bibinfo
  {author} {\bibfnamefont {Z.}~\bibnamefont {Cheng}}, \bibinfo {author}
  {\bibfnamefont {M.}~\bibnamefont {Zhu}}, \bibinfo {author} {\bibfnamefont
  {H.}~\bibnamefont {Zhou}}, \bibinfo {author} {\bibfnamefont {P.}~\bibnamefont
  {Jiang}}, \bibinfo {author} {\bibfnamefont {L.}~\bibnamefont {Zhou}},
  \bibinfo {author} {\bibfnamefont {Q.}~\bibnamefont {Xue}}, \bibinfo {author}
  {\bibfnamefont {F.}~\bibnamefont {Yuan}}, \bibinfo {author} {\bibfnamefont
  {J.}~\bibnamefont {Zhu}},  \emph {et~al.},\ }\href@noop {} {\bibfield
  {journal} {\bibinfo  {journal} {Nature}\ }\textbf {\bibinfo {volume} {592}},\
  \bibinfo {pages} {712} (\bibinfo {year} {2021})}\BibitemShut {NoStop}%
\bibitem [{\citenamefont {Walsh}\ \emph {et~al.}(2021)\citenamefont {Walsh},
  \citenamefont {Asta},\ and\ \citenamefont {Ritchie}}]{walsh2021magnetically}%
  \BibitemOpen
  \bibfield  {author} {\bibinfo {author} {\bibfnamefont {F.}~\bibnamefont
  {Walsh}}, \bibinfo {author} {\bibfnamefont {M.}~\bibnamefont {Asta}}, \ and\
  \bibinfo {author} {\bibfnamefont {R.~O.}\ \bibnamefont {Ritchie}},\
  }\href@noop {} {\bibfield  {journal} {\bibinfo  {journal} {Proceedings of the
  National Academy of Sciences}\ }\textbf {\bibinfo {volume} {118}},\ \bibinfo
  {pages} {e2020540118} (\bibinfo {year} {2021})}\BibitemShut {NoStop}%
\bibitem [{\citenamefont {Zhang}\ \emph {et~al.}(2017)\citenamefont {Zhang},
  \citenamefont {Zhao}, \citenamefont {Jin}, \citenamefont {Xue}, \citenamefont
  {Velisa}, \citenamefont {Bei}, \citenamefont {Huang}, \citenamefont {Ko},
  \citenamefont {Pagan}, \citenamefont {Neuefeind} \emph
  {et~al.}}]{zhang2017local}%
  \BibitemOpen
  \bibfield  {author} {\bibinfo {author} {\bibfnamefont {F.}~\bibnamefont
  {Zhang}}, \bibinfo {author} {\bibfnamefont {S.}~\bibnamefont {Zhao}},
  \bibinfo {author} {\bibfnamefont {K.}~\bibnamefont {Jin}}, \bibinfo {author}
  {\bibfnamefont {H.}~\bibnamefont {Xue}}, \bibinfo {author} {\bibfnamefont
  {G.}~\bibnamefont {Velisa}}, \bibinfo {author} {\bibfnamefont
  {H.}~\bibnamefont {Bei}}, \bibinfo {author} {\bibfnamefont {R.}~\bibnamefont
  {Huang}}, \bibinfo {author} {\bibfnamefont {J.}~\bibnamefont {Ko}}, \bibinfo
  {author} {\bibfnamefont {D.}~\bibnamefont {Pagan}}, \bibinfo {author}
  {\bibfnamefont {J.}~\bibnamefont {Neuefeind}},  \emph {et~al.},\ }\href@noop
  {} {\bibfield  {journal} {\bibinfo  {journal} {Physical review letters}\
  }\textbf {\bibinfo {volume} {118}},\ \bibinfo {pages} {205501} (\bibinfo
  {year} {2017})}\BibitemShut {NoStop}%
\bibitem [{\citenamefont {Ding}\ \emph {et~al.}(2019)\citenamefont {Ding},
  \citenamefont {Zhang}, \citenamefont {Chen}, \citenamefont {Fu},
  \citenamefont {Chen}, \citenamefont {Chen}, \citenamefont {Gu}, \citenamefont
  {Wei}, \citenamefont {Bei}, \citenamefont {Gao}, \citenamefont {Wen},
  \citenamefont {Li}, \citenamefont {Zhang}, \citenamefont {Zhu}, \citenamefont
  {Ritchie},\ and\ \citenamefont {Yu}}]{Ding2019}%
  \BibitemOpen
  \bibfield  {author} {\bibinfo {author} {\bibfnamefont {Q.}~\bibnamefont
  {Ding}}, \bibinfo {author} {\bibfnamefont {Y.}~\bibnamefont {Zhang}},
  \bibinfo {author} {\bibfnamefont {X.}~\bibnamefont {Chen}}, \bibinfo {author}
  {\bibfnamefont {X.}~\bibnamefont {Fu}}, \bibinfo {author} {\bibfnamefont
  {D.}~\bibnamefont {Chen}}, \bibinfo {author} {\bibfnamefont {S.}~\bibnamefont
  {Chen}}, \bibinfo {author} {\bibfnamefont {L.}~\bibnamefont {Gu}}, \bibinfo
  {author} {\bibfnamefont {F.}~\bibnamefont {Wei}}, \bibinfo {author}
  {\bibfnamefont {H.}~\bibnamefont {Bei}}, \bibinfo {author} {\bibfnamefont
  {Y.}~\bibnamefont {Gao}}, \bibinfo {author} {\bibfnamefont {M.}~\bibnamefont
  {Wen}}, \bibinfo {author} {\bibfnamefont {J.}~\bibnamefont {Li}}, \bibinfo
  {author} {\bibfnamefont {Z.}~\bibnamefont {Zhang}}, \bibinfo {author}
  {\bibfnamefont {T.}~\bibnamefont {Zhu}}, \bibinfo {author} {\bibfnamefont
  {R.~O.}\ \bibnamefont {Ritchie}}, \ and\ \bibinfo {author} {\bibfnamefont
  {Q.}~\bibnamefont {Yu}},\ }\href {\doibase 10.1038/s41586-019-1617-1}
  {\bibfield  {journal} {\bibinfo  {journal} {Nature}\ }\textbf {\bibinfo
  {volume} {574}},\ \bibinfo {pages} {223} (\bibinfo {year}
  {2019})}\BibitemShut {NoStop}%
\bibitem [{\citenamefont {Widom}\ \emph {et~al.}(2014)\citenamefont {Widom},
  \citenamefont {Huhn}, \citenamefont {Maiti},\ and\ \citenamefont
  {Steurer}}]{Widom2014}%
  \BibitemOpen
  \bibfield  {author} {\bibinfo {author} {\bibfnamefont {M.}~\bibnamefont
  {Widom}}, \bibinfo {author} {\bibfnamefont {W.~P.}\ \bibnamefont {Huhn}},
  \bibinfo {author} {\bibfnamefont {S.}~\bibnamefont {Maiti}}, \ and\ \bibinfo
  {author} {\bibfnamefont {W.}~\bibnamefont {Steurer}},\ }\href {\doibase
  10.1007/s11661-013-2000-8} {\bibfield  {journal} {\bibinfo  {journal}
  {Metallurgical and Materials Transactions A}\ }\textbf {\bibinfo {volume}
  {45}},\ \bibinfo {pages} {196} (\bibinfo {year} {2014})}\BibitemShut
  {NoStop}%
\bibitem [{\citenamefont {Tamm}\ \emph {et~al.}(2015)\citenamefont {Tamm},
  \citenamefont {Aabloo}, \citenamefont {Klintenberg}, \citenamefont {Stocks},\
  and\ \citenamefont {Caro}}]{TAMM2015307}%
  \BibitemOpen
  \bibfield  {author} {\bibinfo {author} {\bibfnamefont {A.}~\bibnamefont
  {Tamm}}, \bibinfo {author} {\bibfnamefont {A.}~\bibnamefont {Aabloo}},
  \bibinfo {author} {\bibfnamefont {M.}~\bibnamefont {Klintenberg}}, \bibinfo
  {author} {\bibfnamefont {M.}~\bibnamefont {Stocks}}, \ and\ \bibinfo {author}
  {\bibfnamefont {A.}~\bibnamefont {Caro}},\ }\href {\doibase
  https://doi.org/10.1016/j.actamat.2015.08.015} {\bibfield  {journal}
  {\bibinfo  {journal} {Acta Materialia}\ }\textbf {\bibinfo {volume} {99}},\
  \bibinfo {pages} {307} (\bibinfo {year} {2015})}\BibitemShut {NoStop}%
\bibitem [{\citenamefont {Santodonato}\ \emph {et~al.}(2015)\citenamefont
  {Santodonato}, \citenamefont {Zhang}, \citenamefont {Feygenson},
  \citenamefont {Parish}, \citenamefont {Gao}, \citenamefont {Weber},
  \citenamefont {Neuefeind}, \citenamefont {Tang},\ and\ \citenamefont
  {Liaw}}]{Santodonato2015}%
  \BibitemOpen
  \bibfield  {author} {\bibinfo {author} {\bibfnamefont {L.~J.}\ \bibnamefont
  {Santodonato}}, \bibinfo {author} {\bibfnamefont {Y.}~\bibnamefont {Zhang}},
  \bibinfo {author} {\bibfnamefont {M.}~\bibnamefont {Feygenson}}, \bibinfo
  {author} {\bibfnamefont {C.~M.}\ \bibnamefont {Parish}}, \bibinfo {author}
  {\bibfnamefont {M.~C.}\ \bibnamefont {Gao}}, \bibinfo {author} {\bibfnamefont
  {R.~J.}\ \bibnamefont {Weber}}, \bibinfo {author} {\bibfnamefont {J.~C.}\
  \bibnamefont {Neuefeind}}, \bibinfo {author} {\bibfnamefont {Z.}~\bibnamefont
  {Tang}}, \ and\ \bibinfo {author} {\bibfnamefont {P.~K.}\ \bibnamefont
  {Liaw}},\ }\href {\doibase 10.1038/ncomms6964} {\bibfield  {journal}
  {\bibinfo  {journal} {Nature Communications}\ }\textbf {\bibinfo {volume}
  {6}},\ \bibinfo {pages} {5964} (\bibinfo {year} {2015})}\BibitemShut
  {NoStop}%
\bibitem [{\citenamefont {Singh}\ \emph {et~al.}(2015)\citenamefont {Singh},
  \citenamefont {Smirnov},\ and\ \citenamefont {Johnson}}]{singh}%
  \BibitemOpen
  \bibfield  {author} {\bibinfo {author} {\bibfnamefont {P.}~\bibnamefont
  {Singh}}, \bibinfo {author} {\bibfnamefont {A.~V.}\ \bibnamefont {Smirnov}},
  \ and\ \bibinfo {author} {\bibfnamefont {D.~D.}\ \bibnamefont {Johnson}},\
  }\href {\doibase 10.1103/PhysRevB.91.224204} {\bibfield  {journal} {\bibinfo
  {journal} {Phys. Rev. B}\ }\textbf {\bibinfo {volume} {91}},\ \bibinfo
  {pages} {224204} (\bibinfo {year} {2015})}\BibitemShut {NoStop}%
\bibitem [{\citenamefont {Koch}\ \emph {et~al.}(2017)\citenamefont {Koch},
  \citenamefont {Granberg}, \citenamefont {Brink}, \citenamefont {Utt},
  \citenamefont {Albe}, \citenamefont {Djurabekova},\ and\ \citenamefont
  {Nordlund}}]{koch}%
  \BibitemOpen
  \bibfield  {author} {\bibinfo {author} {\bibfnamefont {L.}~\bibnamefont
  {Koch}}, \bibinfo {author} {\bibfnamefont {F.}~\bibnamefont {Granberg}},
  \bibinfo {author} {\bibfnamefont {T.}~\bibnamefont {Brink}}, \bibinfo
  {author} {\bibfnamefont {D.}~\bibnamefont {Utt}}, \bibinfo {author}
  {\bibfnamefont {K.}~\bibnamefont {Albe}}, \bibinfo {author} {\bibfnamefont
  {F.}~\bibnamefont {Djurabekova}}, \ and\ \bibinfo {author} {\bibfnamefont
  {K.}~\bibnamefont {Nordlund}},\ }\href {\doibase 10.1063/1.4990950}
  {\bibfield  {journal} {\bibinfo  {journal} {Journal of Applied Physics}\
  }\textbf {\bibinfo {volume} {122}},\ \bibinfo {pages} {105106} (\bibinfo
  {year} {2017})},\ \Eprint
  {http://arxiv.org/abs/https://doi.org/10.1063/1.4990950}
  {https://doi.org/10.1063/1.4990950} \BibitemShut {NoStop}%
\bibitem [{\citenamefont {Fern{\'a}ndez-Caballero}\ \emph
  {et~al.}(2017)\citenamefont {Fern{\'a}ndez-Caballero}, \citenamefont
  {Wr{\'o}bel}, \citenamefont {Mummery},\ and\ \citenamefont
  {Nguyen-Manh}}]{Fernandez-Caballero2017}%
  \BibitemOpen
  \bibfield  {author} {\bibinfo {author} {\bibfnamefont {A.}~\bibnamefont
  {Fern{\'a}ndez-Caballero}}, \bibinfo {author} {\bibfnamefont {J.~S.}\
  \bibnamefont {Wr{\'o}bel}}, \bibinfo {author} {\bibfnamefont {P.~M.}\
  \bibnamefont {Mummery}}, \ and\ \bibinfo {author} {\bibfnamefont
  {D.}~\bibnamefont {Nguyen-Manh}},\ }\href {\doibase
  10.1007/s11669-017-0582-3} {\bibfield  {journal} {\bibinfo  {journal}
  {Journal of Phase Equilibria and Diffusion}\ }\textbf {\bibinfo {volume}
  {38}},\ \bibinfo {pages} {391} (\bibinfo {year} {2017})}\BibitemShut
  {NoStop}%
\bibitem [{\citenamefont {Ma}\ \emph {et~al.}(2018)\citenamefont {Ma},
  \citenamefont {Wang}, \citenamefont {Li}, \citenamefont {Santodonato},
  \citenamefont {Feygenson}, \citenamefont {Dong},\ and\ \citenamefont
  {Liaw}}]{MA201864}%
  \BibitemOpen
  \bibfield  {author} {\bibinfo {author} {\bibfnamefont {Y.}~\bibnamefont
  {Ma}}, \bibinfo {author} {\bibfnamefont {Q.}~\bibnamefont {Wang}}, \bibinfo
  {author} {\bibfnamefont {C.}~\bibnamefont {Li}}, \bibinfo {author}
  {\bibfnamefont {L.~J.}\ \bibnamefont {Santodonato}}, \bibinfo {author}
  {\bibfnamefont {M.}~\bibnamefont {Feygenson}}, \bibinfo {author}
  {\bibfnamefont {C.}~\bibnamefont {Dong}}, \ and\ \bibinfo {author}
  {\bibfnamefont {P.~K.}\ \bibnamefont {Liaw}},\ }\href {\doibase
  https://doi.org/10.1016/j.scriptamat.2017.09.049} {\bibfield  {journal}
  {\bibinfo  {journal} {Scripta Materialia}\ }\textbf {\bibinfo {volume}
  {144}},\ \bibinfo {pages} {64} (\bibinfo {year} {2018})}\BibitemShut
  {NoStop}%
\bibitem [{\citenamefont {Li}\ \emph {et~al.}(2019{\natexlab{b}})\citenamefont
  {Li}, \citenamefont {Sheng},\ and\ \citenamefont {Ma}}]{Li2019}%
  \BibitemOpen
  \bibfield  {author} {\bibinfo {author} {\bibfnamefont {Q.-J.}\ \bibnamefont
  {Li}}, \bibinfo {author} {\bibfnamefont {H.}~\bibnamefont {Sheng}}, \ and\
  \bibinfo {author} {\bibfnamefont {E.}~\bibnamefont {Ma}},\ }\href {\doibase
  10.1038/s41467-019-11464-7} {\bibfield  {journal} {\bibinfo  {journal}
  {Nature Communications}\ }\textbf {\bibinfo {volume} {10}},\ \bibinfo {pages}
  {3563} (\bibinfo {year} {2019}{\natexlab{b}})}\BibitemShut {NoStop}%
\bibitem [{\citenamefont {Oh}\ \emph {et~al.}(2019)\citenamefont {Oh},
  \citenamefont {Kim}, \citenamefont {Odbadrakh}, \citenamefont {Ryu},
  \citenamefont {Yoon}, \citenamefont {Mu}, \citenamefont {K{\"o}rmann},
  \citenamefont {Ikeda}, \citenamefont {Tasan}, \citenamefont {Raabe},
  \citenamefont {Egami},\ and\ \citenamefont {Park}}]{Oh2019}%
  \BibitemOpen
  \bibfield  {author} {\bibinfo {author} {\bibfnamefont {H.~S.}\ \bibnamefont
  {Oh}}, \bibinfo {author} {\bibfnamefont {S.~J.}\ \bibnamefont {Kim}},
  \bibinfo {author} {\bibfnamefont {K.}~\bibnamefont {Odbadrakh}}, \bibinfo
  {author} {\bibfnamefont {W.~H.}\ \bibnamefont {Ryu}}, \bibinfo {author}
  {\bibfnamefont {K.~N.}\ \bibnamefont {Yoon}}, \bibinfo {author}
  {\bibfnamefont {S.}~\bibnamefont {Mu}}, \bibinfo {author} {\bibfnamefont
  {F.}~\bibnamefont {K{\"o}rmann}}, \bibinfo {author} {\bibfnamefont
  {Y.}~\bibnamefont {Ikeda}}, \bibinfo {author} {\bibfnamefont {C.~C.}\
  \bibnamefont {Tasan}}, \bibinfo {author} {\bibfnamefont {D.}~\bibnamefont
  {Raabe}}, \bibinfo {author} {\bibfnamefont {T.}~\bibnamefont {Egami}}, \ and\
  \bibinfo {author} {\bibfnamefont {E.~S.}\ \bibnamefont {Park}},\ }\href
  {\doibase 10.1038/s41467-019-10012-7} {\bibfield  {journal} {\bibinfo
  {journal} {Nature Communications}\ }\textbf {\bibinfo {volume} {10}},\
  \bibinfo {pages} {2090} (\bibinfo {year} {2019})}\BibitemShut {NoStop}%
\bibitem [{\citenamefont {Jian}\ \emph {et~al.}(2020)\citenamefont {Jian},
  \citenamefont {Xie}, \citenamefont {Xu}, \citenamefont {Su}, \citenamefont
  {Yao},\ and\ \citenamefont {Beyerlein}}]{JIAN2020352}%
  \BibitemOpen
  \bibfield  {author} {\bibinfo {author} {\bibfnamefont {W.-R.}\ \bibnamefont
  {Jian}}, \bibinfo {author} {\bibfnamefont {Z.}~\bibnamefont {Xie}}, \bibinfo
  {author} {\bibfnamefont {S.}~\bibnamefont {Xu}}, \bibinfo {author}
  {\bibfnamefont {Y.}~\bibnamefont {Su}}, \bibinfo {author} {\bibfnamefont
  {X.}~\bibnamefont {Yao}}, \ and\ \bibinfo {author} {\bibfnamefont {I.~J.}\
  \bibnamefont {Beyerlein}},\ }\href {\doibase
  https://doi.org/10.1016/j.actamat.2020.08.044} {\bibfield  {journal}
  {\bibinfo  {journal} {Acta Materialia}\ }\textbf {\bibinfo {volume} {199}},\
  \bibinfo {pages} {352} (\bibinfo {year} {2020})}\BibitemShut {NoStop}%
\bibitem [{\citenamefont {Yin}\ \emph {et~al.}(2020)\citenamefont {Yin},
  \citenamefont {Yoshida}, \citenamefont {Tsuji},\ and\ \citenamefont
  {Curtin}}]{Yin2020}%
  \BibitemOpen
  \bibfield  {author} {\bibinfo {author} {\bibfnamefont {B.}~\bibnamefont
  {Yin}}, \bibinfo {author} {\bibfnamefont {S.}~\bibnamefont {Yoshida}},
  \bibinfo {author} {\bibfnamefont {N.}~\bibnamefont {Tsuji}}, \ and\ \bibinfo
  {author} {\bibfnamefont {W.~A.}\ \bibnamefont {Curtin}},\ }\href {\doibase
  10.1038/s41467-020-16083-1} {\bibfield  {journal} {\bibinfo  {journal}
  {Nature Communications}\ }\textbf {\bibinfo {volume} {11}},\ \bibinfo {pages}
  {2507} (\bibinfo {year} {2020})}\BibitemShut {NoStop}%
\bibitem [{\citenamefont {George}\ \emph {et~al.}(2020)\citenamefont {George},
  \citenamefont {Curtin},\ and\ \citenamefont {Tasan}}]{GEORGE2020435}%
  \BibitemOpen
  \bibfield  {author} {\bibinfo {author} {\bibfnamefont {E.}~\bibnamefont
  {George}}, \bibinfo {author} {\bibfnamefont {W.}~\bibnamefont {Curtin}}, \
  and\ \bibinfo {author} {\bibfnamefont {C.}~\bibnamefont {Tasan}},\ }\href
  {\doibase https://doi.org/10.1016/j.actamat.2019.12.015} {\bibfield
  {journal} {\bibinfo  {journal} {Acta Materialia}\ }\textbf {\bibinfo {volume}
  {188}},\ \bibinfo {pages} {435} (\bibinfo {year} {2020})}\BibitemShut
  {NoStop}%
\bibitem [{\citenamefont {Ding}\ \emph {et~al.}(2018)\citenamefont {Ding},
  \citenamefont {Yu}, \citenamefont {Asta},\ and\ \citenamefont
  {Ritchie}}]{Ding}%
  \BibitemOpen
  \bibfield  {author} {\bibinfo {author} {\bibfnamefont {J.}~\bibnamefont
  {Ding}}, \bibinfo {author} {\bibfnamefont {Q.}~\bibnamefont {Yu}}, \bibinfo
  {author} {\bibfnamefont {M.}~\bibnamefont {Asta}}, \ and\ \bibinfo {author}
  {\bibfnamefont {R.~O.}\ \bibnamefont {Ritchie}},\ }\href {\doibase
  10.1073/pnas.1808660115} {\bibfield  {journal} {\bibinfo  {journal}
  {Proceedings of the National Academy of Sciences}\ }\textbf {\bibinfo
  {volume} {115}},\ \bibinfo {pages} {8919} (\bibinfo {year} {2018})},\ \Eprint
  {http://arxiv.org/abs/https://www.pnas.org/doi/pdf/10.1073/pnas.1808660115}
  {https://www.pnas.org/doi/pdf/10.1073/pnas.1808660115} \BibitemShut {NoStop}%
\bibitem [{\citenamefont {Lenz}\ \emph {et~al.}(2019)\citenamefont {Lenz},
  \citenamefont {Eggeler}, \citenamefont {M{\"u}ller}, \citenamefont {Zenk},
  \citenamefont {Volz}, \citenamefont {Wollgramm}, \citenamefont {Eggeler},
  \citenamefont {Neumeier}, \citenamefont {G{\"o}ken},\ and\ \citenamefont
  {Spiecker}}]{Lenz2019TensionCompressionAO}%
  \BibitemOpen
  \bibfield  {author} {\bibinfo {author} {\bibfnamefont {M.}~\bibnamefont
  {Lenz}}, \bibinfo {author} {\bibfnamefont {Y.~M.}\ \bibnamefont {Eggeler}},
  \bibinfo {author} {\bibfnamefont {J.~G.}\ \bibnamefont {M{\"u}ller}},
  \bibinfo {author} {\bibfnamefont {C.~H.}\ \bibnamefont {Zenk}}, \bibinfo
  {author} {\bibfnamefont {N.}~\bibnamefont {Volz}}, \bibinfo {author}
  {\bibfnamefont {P.}~\bibnamefont {Wollgramm}}, \bibinfo {author}
  {\bibfnamefont {G.}~\bibnamefont {Eggeler}}, \bibinfo {author} {\bibfnamefont
  {S.}~\bibnamefont {Neumeier}}, \bibinfo {author} {\bibfnamefont
  {M.}~\bibnamefont {G{\"o}ken}}, \ and\ \bibinfo {author} {\bibfnamefont
  {E.}~\bibnamefont {Spiecker}},\ }\href@noop {} {\bibfield  {journal}
  {\bibinfo  {journal} {Acta Materialia}\ } (\bibinfo {year}
  {2019})}\BibitemShut {NoStop}%
\bibitem [{\citenamefont {Bezold}\ \emph {et~al.}(2020)\citenamefont {Bezold},
  \citenamefont {Volz}, \citenamefont {Xue}, \citenamefont {Zenk},
  \citenamefont {Neumeier},\ and\ \citenamefont {G{\"o}ken}}]{Bezold2020}%
  \BibitemOpen
  \bibfield  {author} {\bibinfo {author} {\bibfnamefont {A.}~\bibnamefont
  {Bezold}}, \bibinfo {author} {\bibfnamefont {N.}~\bibnamefont {Volz}},
  \bibinfo {author} {\bibfnamefont {F.}~\bibnamefont {Xue}}, \bibinfo {author}
  {\bibfnamefont {C.~H.}\ \bibnamefont {Zenk}}, \bibinfo {author}
  {\bibfnamefont {S.}~\bibnamefont {Neumeier}}, \ and\ \bibinfo {author}
  {\bibfnamefont {M.}~\bibnamefont {G{\"o}ken}},\ }\href {\doibase
  10.1007/s11661-020-05626-2} {\bibfield  {journal} {\bibinfo  {journal}
  {Metallurgical and Materials Transactions A}\ }\textbf {\bibinfo {volume}
  {51}},\ \bibinfo {pages} {1567} (\bibinfo {year} {2020})}\BibitemShut
  {NoStop}%
\bibitem [{\citenamefont {Xue}\ \emph {et~al.}(2018)\citenamefont {Xue},
  \citenamefont {Zenk}, \citenamefont {Freund}, \citenamefont {Hoelzel},
  \citenamefont {Neumeier},\ and\ \citenamefont {Göken}}]{XUE2018129}%
  \BibitemOpen
  \bibfield  {author} {\bibinfo {author} {\bibfnamefont {F.}~\bibnamefont
  {Xue}}, \bibinfo {author} {\bibfnamefont {C.}~\bibnamefont {Zenk}}, \bibinfo
  {author} {\bibfnamefont {L.}~\bibnamefont {Freund}}, \bibinfo {author}
  {\bibfnamefont {M.}~\bibnamefont {Hoelzel}}, \bibinfo {author} {\bibfnamefont
  {S.}~\bibnamefont {Neumeier}}, \ and\ \bibinfo {author} {\bibfnamefont
  {M.}~\bibnamefont {Göken}},\ }\href {\doibase
  https://doi.org/10.1016/j.scriptamat.2017.08.039} {\bibfield  {journal}
  {\bibinfo  {journal} {Scripta Materialia}\ }\textbf {\bibinfo {volume}
  {142}},\ \bibinfo {pages} {129} (\bibinfo {year} {2018})}\BibitemShut
  {NoStop}%
\bibitem [{\citenamefont {Zenk}\ \emph {et~al.}(2020)\citenamefont {Zenk},
  \citenamefont {Volz}, \citenamefont {Zenk}, \citenamefont {Felfer},\ and\
  \citenamefont {Neumeier}}]{cryst10111058}%
  \BibitemOpen
  \bibfield  {author} {\bibinfo {author} {\bibfnamefont {C.~H.}\ \bibnamefont
  {Zenk}}, \bibinfo {author} {\bibfnamefont {N.}~\bibnamefont {Volz}}, \bibinfo
  {author} {\bibfnamefont {C.}~\bibnamefont {Zenk}}, \bibinfo {author}
  {\bibfnamefont {P.~J.}\ \bibnamefont {Felfer}}, \ and\ \bibinfo {author}
  {\bibfnamefont {S.}~\bibnamefont {Neumeier}},\ }\href {\doibase
  10.3390/cryst10111058} {\bibfield  {journal} {\bibinfo  {journal} {Crystals}\
  }\textbf {\bibinfo {volume} {10}} (\bibinfo {year} {2020}),\
  10.3390/cryst10111058}\BibitemShut {NoStop}%
\bibitem [{\citenamefont {Rhein}\ \emph {et~al.}(2018)\citenamefont {Rhein},
  \citenamefont {Callahan}, \citenamefont {Murray}, \citenamefont {Stinville},
  \citenamefont {Titus}, \citenamefont {Van~der Ven},\ and\ \citenamefont
  {Pollock}}]{Rhein2018}%
  \BibitemOpen
  \bibfield  {author} {\bibinfo {author} {\bibfnamefont {R.~K.}\ \bibnamefont
  {Rhein}}, \bibinfo {author} {\bibfnamefont {P.~G.}\ \bibnamefont {Callahan}},
  \bibinfo {author} {\bibfnamefont {S.~P.}\ \bibnamefont {Murray}}, \bibinfo
  {author} {\bibfnamefont {J.-C.}\ \bibnamefont {Stinville}}, \bibinfo {author}
  {\bibfnamefont {M.~S.}\ \bibnamefont {Titus}}, \bibinfo {author}
  {\bibfnamefont {A.}~\bibnamefont {Van~der Ven}}, \ and\ \bibinfo {author}
  {\bibfnamefont {T.~M.}\ \bibnamefont {Pollock}},\ }\href {\doibase
  10.1007/s11661-018-4768-z} {\bibfield  {journal} {\bibinfo  {journal}
  {Metallurgical and Materials Transactions A}\ }\textbf {\bibinfo {volume}
  {49}},\ \bibinfo {pages} {4090} (\bibinfo {year} {2018})}\BibitemShut
  {NoStop}%
\bibitem [{\citenamefont {Feng}\ \emph {et~al.}(2018)\citenamefont {Feng},
  \citenamefont {Lv}, \citenamefont {Rhein}, \citenamefont {Goiri},
  \citenamefont {Titus}, \citenamefont {{Van der Ven}}, \citenamefont
  {Pollock},\ and\ \citenamefont {Wang}}]{FENG201899}%
  \BibitemOpen
  \bibfield  {author} {\bibinfo {author} {\bibfnamefont {L.}~\bibnamefont
  {Feng}}, \bibinfo {author} {\bibfnamefont {D.}~\bibnamefont {Lv}}, \bibinfo
  {author} {\bibfnamefont {R.}~\bibnamefont {Rhein}}, \bibinfo {author}
  {\bibfnamefont {J.}~\bibnamefont {Goiri}}, \bibinfo {author} {\bibfnamefont
  {M.}~\bibnamefont {Titus}}, \bibinfo {author} {\bibfnamefont
  {A.}~\bibnamefont {{Van der Ven}}}, \bibinfo {author} {\bibfnamefont
  {T.}~\bibnamefont {Pollock}}, \ and\ \bibinfo {author} {\bibfnamefont
  {Y.}~\bibnamefont {Wang}},\ }\href {\doibase
  https://doi.org/10.1016/j.actamat.2018.09.013} {\bibfield  {journal}
  {\bibinfo  {journal} {Acta Materialia}\ }\textbf {\bibinfo {volume} {161}},\
  \bibinfo {pages} {99} (\bibinfo {year} {2018})}\BibitemShut {NoStop}%
\bibitem [{\citenamefont {May}\ \emph {et~al.}(2005)\citenamefont {May},
  \citenamefont {H{\"o}ppel},\ and\ \citenamefont {G{\"o}ken}}]{may2005strain}%
  \BibitemOpen
  \bibfield  {author} {\bibinfo {author} {\bibfnamefont {J.}~\bibnamefont
  {May}}, \bibinfo {author} {\bibfnamefont {H.}~\bibnamefont {H{\"o}ppel}}, \
  and\ \bibinfo {author} {\bibfnamefont {M.}~\bibnamefont {G{\"o}ken}},\
  }\href@noop {} {\bibfield  {journal} {\bibinfo  {journal} {Scripta
  Materialia}\ }\textbf {\bibinfo {volume} {53}},\ \bibinfo {pages} {189}
  (\bibinfo {year} {2005})}\BibitemShut {NoStop}%
\bibitem [{\citenamefont {Chinh}\ \emph {et~al.}(2006)\citenamefont {Chinh},
  \citenamefont {Szommer}, \citenamefont {Horita},\ and\ \citenamefont
  {Langdon}}]{Chinh}%
  \BibitemOpen
  \bibfield  {author} {\bibinfo {author} {\bibfnamefont {N.}~\bibnamefont
  {Chinh}}, \bibinfo {author} {\bibfnamefont {P.}~\bibnamefont {Szommer}},
  \bibinfo {author} {\bibfnamefont {Z.}~\bibnamefont {Horita}}, \ and\ \bibinfo
  {author} {\bibfnamefont {T.}~\bibnamefont {Langdon}},\ }\href {\doibase
  https://doi.org/10.1002/adma.200501232} {\bibfield  {journal} {\bibinfo
  {journal} {Advanced Materials}\ }\textbf {\bibinfo {volume} {18}},\ \bibinfo
  {pages} {34} (\bibinfo {year} {2006})},\ \Eprint
  {http://arxiv.org/abs/https://onlinelibrary.wiley.com/doi/pdf/10.1002/adma.200501232}
  {https://onlinelibrary.wiley.com/doi/pdf/10.1002/adma.200501232} \BibitemShut
  {NoStop}%
\bibitem [{\citenamefont {Li}\ \emph {et~al.}(2007)\citenamefont {Li},
  \citenamefont {Mueller}, \citenamefont {Höppel}, \citenamefont {Göken},\
  and\ \citenamefont {Blum}}]{Li2007}%
  \BibitemOpen
  \bibfield  {author} {\bibinfo {author} {\bibfnamefont {Y.}~\bibnamefont
  {Li}}, \bibinfo {author} {\bibfnamefont {J.}~\bibnamefont {Mueller}},
  \bibinfo {author} {\bibfnamefont {H.}~\bibnamefont {Höppel}}, \bibinfo
  {author} {\bibfnamefont {M.}~\bibnamefont {Göken}}, \ and\ \bibinfo {author}
  {\bibfnamefont {W.}~\bibnamefont {Blum}},\ }\href {\doibase
  10.1016/j.actamat.2007.06.036} {\bibfield  {journal} {\bibinfo  {journal}
  {Acta Materialia}\ }\textbf {\bibinfo {volume} {55}},\ \bibinfo {pages}
  {5708} (\bibinfo {year} {2007})}\BibitemShut {NoStop}%
\bibitem [{\citenamefont {Wei}\ \emph {et~al.}(2004)\citenamefont {Wei},
  \citenamefont {Cheng}, \citenamefont {Ramesh},\ and\ \citenamefont
  {Ma}}]{WEI200471}%
  \BibitemOpen
  \bibfield  {author} {\bibinfo {author} {\bibfnamefont {Q.}~\bibnamefont
  {Wei}}, \bibinfo {author} {\bibfnamefont {S.}~\bibnamefont {Cheng}}, \bibinfo
  {author} {\bibfnamefont {K.}~\bibnamefont {Ramesh}}, \ and\ \bibinfo {author}
  {\bibfnamefont {E.}~\bibnamefont {Ma}},\ }\href {\doibase
  https://doi.org/10.1016/j.msea.2004.03.064} {\bibfield  {journal} {\bibinfo
  {journal} {Materials Science and Engineering: A}\ }\textbf {\bibinfo {volume}
  {381}},\ \bibinfo {pages} {71} (\bibinfo {year} {2004})}\BibitemShut
  {NoStop}%
\bibitem [{\citenamefont {Höppel}\ \emph {et~al.}(2004)\citenamefont
  {Höppel}, \citenamefont {May},\ and\ \citenamefont {Göken}}]{Hoppel}%
  \BibitemOpen
  \bibfield  {author} {\bibinfo {author} {\bibfnamefont {H.}~\bibnamefont
  {Höppel}}, \bibinfo {author} {\bibfnamefont {J.}~\bibnamefont {May}}, \ and\
  \bibinfo {author} {\bibfnamefont {M.}~\bibnamefont {Göken}},\ }\href
  {\doibase https://doi.org/10.1002/adem.200306582} {\bibfield  {journal}
  {\bibinfo  {journal} {Advanced Engineering Materials}\ }\textbf {\bibinfo
  {volume} {6}},\ \bibinfo {pages} {781} (\bibinfo {year} {2004})},\ \Eprint
  {http://arxiv.org/abs/https://onlinelibrary.wiley.com/doi/pdf/10.1002/adem.200306582}
  {https://onlinelibrary.wiley.com/doi/pdf/10.1002/adem.200306582} \BibitemShut
  {NoStop}%
\bibitem [{\citenamefont {Ni}\ \emph {et~al.}(2017)\citenamefont {Ni},
  \citenamefont {Papanikolaou}, \citenamefont {Vajente}, \citenamefont
  {Adhikari},\ and\ \citenamefont {Greer}}]{ni2017probing}%
  \BibitemOpen
  \bibfield  {author} {\bibinfo {author} {\bibfnamefont {X.}~\bibnamefont
  {Ni}}, \bibinfo {author} {\bibfnamefont {S.}~\bibnamefont {Papanikolaou}},
  \bibinfo {author} {\bibfnamefont {G.}~\bibnamefont {Vajente}}, \bibinfo
  {author} {\bibfnamefont {R.~X.}\ \bibnamefont {Adhikari}}, \ and\ \bibinfo
  {author} {\bibfnamefont {J.~R.}\ \bibnamefont {Greer}},\ }\href@noop {}
  {\bibfield  {journal} {\bibinfo  {journal} {Physical review letters}\
  }\textbf {\bibinfo {volume} {118}},\ \bibinfo {pages} {155501} (\bibinfo
  {year} {2017})}\BibitemShut {NoStop}%
\bibitem [{\citenamefont {Maier}\ \emph {et~al.}(2013)\citenamefont {Maier},
  \citenamefont {Merle}, \citenamefont {G{\"o}ken},\ and\ \citenamefont
  {Durst}}]{Maier2013}%
  \BibitemOpen
  \bibfield  {author} {\bibinfo {author} {\bibfnamefont {V.}~\bibnamefont
  {Maier}}, \bibinfo {author} {\bibfnamefont {B.}~\bibnamefont {Merle}},
  \bibinfo {author} {\bibfnamefont {M.}~\bibnamefont {G{\"o}ken}}, \ and\
  \bibinfo {author} {\bibfnamefont {K.}~\bibnamefont {Durst}},\ }\href
  {\doibase 10.1557/jmr.2013.39} {\bibfield  {journal} {\bibinfo  {journal}
  {Journal of Materials Research}\ }\textbf {\bibinfo {volume} {28}},\ \bibinfo
  {pages} {1177} (\bibinfo {year} {2013})}\BibitemShut {NoStop}%
\bibitem [{\citenamefont {Poisl}\ \emph {et~al.}(1995)\citenamefont {Poisl},
  \citenamefont {Oliver},\ and\ \citenamefont {Fabes}}]{Poisl1995}%
  \BibitemOpen
  \bibfield  {author} {\bibinfo {author} {\bibfnamefont {W.~H.}\ \bibnamefont
  {Poisl}}, \bibinfo {author} {\bibfnamefont {W.~C.}\ \bibnamefont {Oliver}}, \
  and\ \bibinfo {author} {\bibfnamefont {B.~D.}\ \bibnamefont {Fabes}},\ }\href
  {\doibase 10.1557/JMR.1995.2024} {\bibfield  {journal} {\bibinfo  {journal}
  {Journal of Materials Research}\ }\textbf {\bibinfo {volume} {10}},\ \bibinfo
  {pages} {2024} (\bibinfo {year} {1995})}\BibitemShut {NoStop}%
\bibitem [{\citenamefont {Lucas}\ and\ \citenamefont
  {Oliver}(1999)}]{Lucas1999}%
  \BibitemOpen
  \bibfield  {author} {\bibinfo {author} {\bibfnamefont {B.~N.}\ \bibnamefont
  {Lucas}}\ and\ \bibinfo {author} {\bibfnamefont {W.~C.}\ \bibnamefont
  {Oliver}},\ }\href {\doibase 10.1007/s11661-999-0051-7} {\bibfield  {journal}
  {\bibinfo  {journal} {Metallurgical and Materials Transactions A}\ }\textbf
  {\bibinfo {volume} {30}},\ \bibinfo {pages} {601} (\bibinfo {year}
  {1999})}\BibitemShut {NoStop}%
\bibitem [{\citenamefont {Stone}\ \emph {et~al.}(2010)\citenamefont {Stone},
  \citenamefont {Joseph}, \citenamefont {Puthoff},\ and\ \citenamefont
  {Elmustafa}}]{Stone2010}%
  \BibitemOpen
  \bibfield  {author} {\bibinfo {author} {\bibfnamefont {D.~S.}\ \bibnamefont
  {Stone}}, \bibinfo {author} {\bibfnamefont {J.~E.}\ \bibnamefont {Joseph}},
  \bibinfo {author} {\bibfnamefont {J.}~\bibnamefont {Puthoff}}, \ and\
  \bibinfo {author} {\bibfnamefont {A.~A.}\ \bibnamefont {Elmustafa}},\ }\href
  {\doibase 10.1557/JMR.2010.0092} {\bibfield  {journal} {\bibinfo  {journal}
  {Journal of Materials Research}\ }\textbf {\bibinfo {volume} {25}},\ \bibinfo
  {pages} {611} (\bibinfo {year} {2010})}\BibitemShut {NoStop}%
\bibitem [{\citenamefont {Choi}\ \emph {et~al.}(2012)\citenamefont {Choi},
  \citenamefont {Yoo}, \citenamefont {Kim},\ and\ \citenamefont
  {Jang}}]{Choi2012}%
  \BibitemOpen
  \bibfield  {author} {\bibinfo {author} {\bibfnamefont {I.-C.}\ \bibnamefont
  {Choi}}, \bibinfo {author} {\bibfnamefont {B.-G.}\ \bibnamefont {Yoo}},
  \bibinfo {author} {\bibfnamefont {Y.-J.}\ \bibnamefont {Kim}}, \ and\
  \bibinfo {author} {\bibfnamefont {J.-i.}\ \bibnamefont {Jang}},\ }\href
  {\doibase 10.1557/jmr.2011.213} {\bibfield  {journal} {\bibinfo  {journal}
  {Journal of Materials Research}\ }\textbf {\bibinfo {volume} {27}},\ \bibinfo
  {pages} {3} (\bibinfo {year} {2012})}\BibitemShut {NoStop}%
\bibitem [{\citenamefont {Thompson}\ \emph {et~al.}(2022)\citenamefont
  {Thompson}, \citenamefont {Aktulga}, \citenamefont {Berger}, \citenamefont
  {Bolintineanu}, \citenamefont {Brown}, \citenamefont {Crozier}, \citenamefont
  {in~'t Veld}, \citenamefont {Kohlmeyer}, \citenamefont {Moore}, \citenamefont
  {Nguyen}, \citenamefont {Shan}, \citenamefont {Stevens}, \citenamefont
  {Tranchida}, \citenamefont {Trott},\ and\ \citenamefont {Plimpton}}]{LAMMPS}%
  \BibitemOpen
  \bibfield  {author} {\bibinfo {author} {\bibfnamefont {A.~P.}\ \bibnamefont
  {Thompson}}, \bibinfo {author} {\bibfnamefont {H.~M.}\ \bibnamefont
  {Aktulga}}, \bibinfo {author} {\bibfnamefont {R.}~\bibnamefont {Berger}},
  \bibinfo {author} {\bibfnamefont {D.~S.}\ \bibnamefont {Bolintineanu}},
  \bibinfo {author} {\bibfnamefont {W.~M.}\ \bibnamefont {Brown}}, \bibinfo
  {author} {\bibfnamefont {P.~S.}\ \bibnamefont {Crozier}}, \bibinfo {author}
  {\bibfnamefont {P.~J.}\ \bibnamefont {in~'t Veld}}, \bibinfo {author}
  {\bibfnamefont {A.}~\bibnamefont {Kohlmeyer}}, \bibinfo {author}
  {\bibfnamefont {S.~G.}\ \bibnamefont {Moore}}, \bibinfo {author}
  {\bibfnamefont {T.~D.}\ \bibnamefont {Nguyen}}, \bibinfo {author}
  {\bibfnamefont {R.}~\bibnamefont {Shan}}, \bibinfo {author} {\bibfnamefont
  {M.~J.}\ \bibnamefont {Stevens}}, \bibinfo {author} {\bibfnamefont
  {J.}~\bibnamefont {Tranchida}}, \bibinfo {author} {\bibfnamefont
  {C.}~\bibnamefont {Trott}}, \ and\ \bibinfo {author} {\bibfnamefont {S.~J.}\
  \bibnamefont {Plimpton}},\ }\href {\doibase 10.1016/j.cpc.2021.108171}
  {\bibfield  {journal} {\bibinfo  {journal} {Comp. Phys. Comm.}\ }\textbf
  {\bibinfo {volume} {271}},\ \bibinfo {pages} {108171} (\bibinfo {year}
  {2022})}\BibitemShut {NoStop}%
\bibitem [{\citenamefont {Domínguez-Gutiérrez}\ \emph
  {et~al.}(2021)\citenamefont {Domínguez-Gutiérrez}, \citenamefont
  {Papanikolaou}, \citenamefont {Esfandiarpour}, \citenamefont {Sobkowicz},\
  and\ \citenamefont {Alava}}]{DOMINGUEZGUTIERREZ2021141912}%
  \BibitemOpen
  \bibfield  {author} {\bibinfo {author} {\bibfnamefont {F.}~\bibnamefont
  {Domínguez-Gutiérrez}}, \bibinfo {author} {\bibfnamefont {S.}~\bibnamefont
  {Papanikolaou}}, \bibinfo {author} {\bibfnamefont {A.}~\bibnamefont
  {Esfandiarpour}}, \bibinfo {author} {\bibfnamefont {P.}~\bibnamefont
  {Sobkowicz}}, \ and\ \bibinfo {author} {\bibfnamefont {M.}~\bibnamefont
  {Alava}},\ }\href {\doibase https://doi.org/10.1016/j.msea.2021.141912}
  {\bibfield  {journal} {\bibinfo  {journal} {Materials Science and
  Engineering: A}\ }\textbf {\bibinfo {volume} {826}},\ \bibinfo {pages}
  {141912} (\bibinfo {year} {2021})}\BibitemShut {NoStop}%
\bibitem [{\citenamefont {Kurpaska}\ \emph {et~al.}(2022)\citenamefont
  {Kurpaska}, \citenamefont {Dominguez-Gutierrez}, \citenamefont {Zhang},
  \citenamefont {Mulewska}, \citenamefont {Bei}, \citenamefont {Weber},
  \citenamefont {KosiĹ„ska}, \citenamefont {Chrominski}, \citenamefont
  {Jozwik}, \citenamefont {Alvarez-Donado}, \citenamefont {Papanikolaou},
  \citenamefont {Jagielski},\ and\ \citenamefont {Alava}}]{KURPASKA2022110639}%
  \BibitemOpen
  \bibfield  {author} {\bibinfo {author} {\bibfnamefont {L.}~\bibnamefont
  {Kurpaska}}, \bibinfo {author} {\bibfnamefont {F.}~\bibnamefont
  {Dominguez-Gutierrez}}, \bibinfo {author} {\bibfnamefont {Y.}~\bibnamefont
  {Zhang}}, \bibinfo {author} {\bibfnamefont {K.}~\bibnamefont {Mulewska}},
  \bibinfo {author} {\bibfnamefont {H.}~\bibnamefont {Bei}}, \bibinfo {author}
  {\bibfnamefont {W.}~\bibnamefont {Weber}}, \bibinfo {author} {\bibfnamefont
  {A.}~\bibnamefont {KosiĹ„ska}}, \bibinfo {author} {\bibfnamefont
  {W.}~\bibnamefont {Chrominski}}, \bibinfo {author} {\bibfnamefont
  {I.}~\bibnamefont {Jozwik}}, \bibinfo {author} {\bibfnamefont
  {R.}~\bibnamefont {Alvarez-Donado}}, \bibinfo {author} {\bibfnamefont
  {S.}~\bibnamefont {Papanikolaou}}, \bibinfo {author} {\bibfnamefont
  {J.}~\bibnamefont {Jagielski}}, \ and\ \bibinfo {author} {\bibfnamefont
  {M.}~\bibnamefont {Alava}},\ }\href {\doibase
  https://doi.org/10.1016/j.matdes.2022.110639} {\bibfield  {journal} {\bibinfo
   {journal} {Materials and Design}\ }\textbf {\bibinfo {volume} {217}},\
  \bibinfo {pages} {110639} (\bibinfo {year} {2022})}\BibitemShut {NoStop}%
\bibitem [{\citenamefont {Sadigh}\ \emph {et~al.}(2012)\citenamefont {Sadigh},
  \citenamefont {Erhart}, \citenamefont {Stukowski}, \citenamefont {Caro},
  \citenamefont {Martinez},\ and\ \citenamefont
  {Zepeda-Ruiz}}]{PhysRevB.85.184203}%
  \BibitemOpen
  \bibfield  {author} {\bibinfo {author} {\bibfnamefont {B.}~\bibnamefont
  {Sadigh}}, \bibinfo {author} {\bibfnamefont {P.}~\bibnamefont {Erhart}},
  \bibinfo {author} {\bibfnamefont {A.}~\bibnamefont {Stukowski}}, \bibinfo
  {author} {\bibfnamefont {A.}~\bibnamefont {Caro}}, \bibinfo {author}
  {\bibfnamefont {E.}~\bibnamefont {Martinez}}, \ and\ \bibinfo {author}
  {\bibfnamefont {L.}~\bibnamefont {Zepeda-Ruiz}},\ }\href {\doibase
  10.1103/PhysRevB.85.184203} {\bibfield  {journal} {\bibinfo  {journal} {Phys.
  Rev. B}\ }\textbf {\bibinfo {volume} {85}},\ \bibinfo {pages} {184203}
  (\bibinfo {year} {2012})}\BibitemShut {NoStop}%
\bibitem [{\citenamefont {Esfandiarpour}\ \emph {et~al.}(2022)\citenamefont
  {Esfandiarpour}, \citenamefont {Papanikolaou},\ and\ \citenamefont
  {Alava}}]{esfandiarpour2022edge}%
  \BibitemOpen
  \bibfield  {author} {\bibinfo {author} {\bibfnamefont {A.}~\bibnamefont
  {Esfandiarpour}}, \bibinfo {author} {\bibfnamefont {S.}~\bibnamefont
  {Papanikolaou}}, \ and\ \bibinfo {author} {\bibfnamefont {M.}~\bibnamefont
  {Alava}},\ }\href@noop {} {\bibfield  {journal} {\bibinfo  {journal}
  {Physical Review Research}\ }\textbf {\bibinfo {volume} {4}},\ \bibinfo
  {pages} {L022043} (\bibinfo {year} {2022})}\BibitemShut {NoStop}%
\bibitem [{\citenamefont {Naghdi}\ \emph {et~al.}(2022)\citenamefont {Naghdi},
  \citenamefont {Karimi}, \citenamefont {Poisvert}, \citenamefont
  {Esfandiarpour}, \citenamefont {Alvarez}, \citenamefont {Sobkowicz},
  \citenamefont {Alava},\ and\ \citenamefont
  {Papanikolaou}}]{naghdi2022dislocation}%
  \BibitemOpen
  \bibfield  {author} {\bibinfo {author} {\bibfnamefont {A.~H.}\ \bibnamefont
  {Naghdi}}, \bibinfo {author} {\bibfnamefont {K.}~\bibnamefont {Karimi}},
  \bibinfo {author} {\bibfnamefont {A.~E.}\ \bibnamefont {Poisvert}}, \bibinfo
  {author} {\bibfnamefont {A.}~\bibnamefont {Esfandiarpour}}, \bibinfo {author}
  {\bibfnamefont {R.}~\bibnamefont {Alvarez}}, \bibinfo {author} {\bibfnamefont
  {P.}~\bibnamefont {Sobkowicz}}, \bibinfo {author} {\bibfnamefont
  {M.}~\bibnamefont {Alava}}, \ and\ \bibinfo {author} {\bibfnamefont
  {S.}~\bibnamefont {Papanikolaou}},\ }\href@noop {} {\bibfield  {journal}
  {\bibinfo  {journal} {arXiv preprint arXiv:2210.12512}\ } (\bibinfo {year}
  {2022})}\BibitemShut {NoStop}%
\bibitem [{\citenamefont {Papanikolaou}\ \emph {et~al.}(2017)\citenamefont
  {Papanikolaou}, \citenamefont {Cui},\ and\ \citenamefont
  {Ghoniem}}]{papanikolaou2017avalanches}%
  \BibitemOpen
  \bibfield  {author} {\bibinfo {author} {\bibfnamefont {S.}~\bibnamefont
  {Papanikolaou}}, \bibinfo {author} {\bibfnamefont {Y.}~\bibnamefont {Cui}}, \
  and\ \bibinfo {author} {\bibfnamefont {N.}~\bibnamefont {Ghoniem}},\
  }\href@noop {} {\bibfield  {journal} {\bibinfo  {journal} {Modelling and
  Simulation in Materials Science and Engineering}\ }\textbf {\bibinfo {volume}
  {26}},\ \bibinfo {pages} {013001} (\bibinfo {year} {2017})}\BibitemShut
  {NoStop}%
\end{thebibliography}

%

\end{document}